\documentclass[preprint,12pt,authoryear]{elsarticle}

\usepackage{amssymb}
\usepackage{amsmath}
\usepackage{bm}
\usepackage{subcaption} 
\usepackage{booktabs} % just makes the tables look nicer
\usepackage{hyperref}

\journal{Spatial Statistics}

\begin{document}

\begin{frontmatter}



\title{A Non-stationary, Amortized, Transfer Learning Approach for Modeling Italian Air Quality} %% Article title


% \author{Alessandro Fusta Moro\fnmark[*], Antony Sikorski\fnmark[*], Daniel McKenzie, Alessandro Fassò, Douglas Nychka} %% Author name

% \fntext[*]{These authors contributed equally.}

% \affiliation{organization={},%Department and Organization

\author[1]{Alessandro Fusta Moro\fnref{equal}\corref{cor1}}
\ead{alessandro.fustamoro@unibg.it}

\author[2]{Antony Sikorski\fnref{equal}}
\author[2]{Daniel McKenzie}
\author[3]{Alessandro Fassò}
\author[2]{Douglas Nychka}

\fntext[equal]{Alessandro Fusta Moro and Antony Sikorski contributed equally to this work.}

\cortext[cor1]{Corresponding author}

\affiliation[1]{organization={University of Bergamo},
            addressline={Department of Engineering and Applied Science},
            city={Bergamo},
            country={Italy}}
            
\affiliation[2]{organization={Colorado School of Mines},
            addressline={Department of Applied Mathematics and Statistics},
            city={Golden},
            state={CO},
            postcode={80401},
            country={USA}}

\affiliation[3]{organization={University of Bergamo},
            addressline={Department of Economics},
            city={Bergamo},
            country={Italy}}

%% Abstract
\begin{abstract}

Air quality monitoring in Italy relies on sparse, irregular, ground-based stations that provide high-quality but incomplete measurements of pollution. Chemical transport models (CTMs) offer full spatial and temporal coverage but smooth over local variability. We develop a spatial transfer-learning framework that integrates these two data sources to produce daily, fine-grid predictions of nitrogen dioxide (NO$_2$) concentrations across Italy for 2023, with uncertainty quantification. The resulting maps provide a resource for decision making in downstream applications such as epidemiology and environmental policy. 

Our approach builds on the geostatistical LatticeKrig framework, which uses compactly supported basis functions and coefficients governed by a sparse precision matrix. We learn a nonstationary, anisotropic correlation structure from the gridded CTM outputs using an image-to-image neural architecture that estimates millions of spatially varying parameters in a matter of seconds. The basis-function representation enables this covariance structure to be transferred to the point-level station data and projected onto a finer prediction grid, a key extension for handling the change of support between data sources. A likelihood-based refinement step then adjusts the correlation range to recover fine-scale variability smoothed out by the gridded data. The proposed methodology results in a flexible, non-stationary, and anisotropic representation of the spatial process, better accommodating the complex geography of Italy. Performance is assessed through experiments on both gridded CTM outputs and point-level station measurements, demonstrating improvements over the stationary formulation.

\end{abstract}

\begin{graphicalabstract}
\includegraphics[width=1\textwidth]{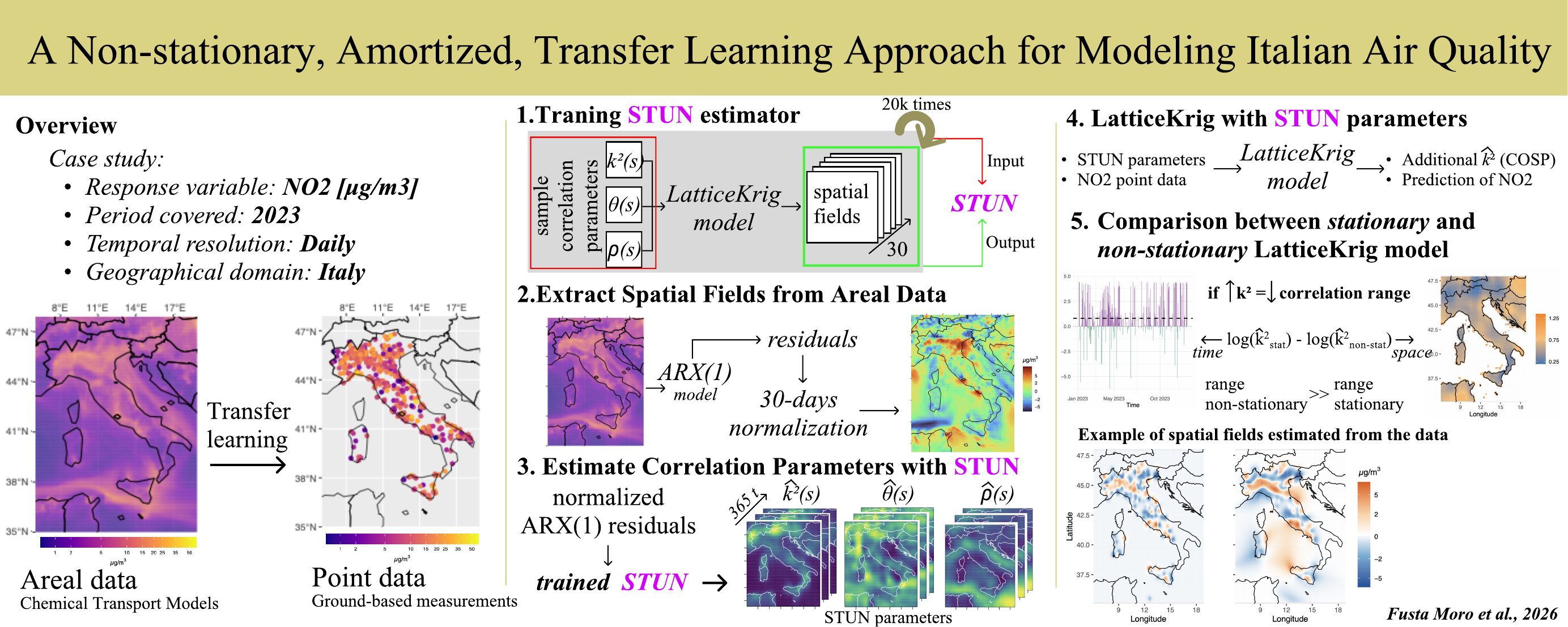}
\end{graphicalabstract}

\begin{highlights}
    \item Amortized neural estimation extended to transfer learning across spatial supports. 
    \item Millions of spatially varying parameters estimated in seconds. 
    \item Parameters adjusted for local effects when moving from gridded to point data. 
    \item Daily, fine-grid NO$_2$ maps of Italy with uncertainties produced for all of 2023.
    \item Non-stationary, anisotropic model better accommodates  complex Italian geography. 
\end{highlights}

%% Keywords

\end{frontmatter}




% remove in final version %%%%%%%%%%%%%%%%%%
%\today

%%%%%%%%%%%%%%%%%%% INTRO %%%%%%%%%%%%%%%%%%%%%
\section{Introduction}
\label{sec1}

Air quality in Italy has been a persistent scientific and political concern due to the high concentrations of atmospheric pollutants regularly detected by Italian and European pollution monitoring systems. Northern Italy is particularly affected: high anthropogenic presence results in emissions from traffic, industrial activity, and livestock farming, which are subsequently trapped by the surrounding mountain ranges \citep{fasso2023agrimonia,otto2024spatiotemporal}. This has prompted the European Union to open repeated infringement proceedings against Italy for not meeting air quality standards. Beyond regulatory considerations, a deeper understanding of air pollutant concentrations is critical to furthering studies in public health \citep{capobussi2016air,conti2018association}, COVID-19 \citep{coker2020effects,gatti2020machine}, and urban planning \citep{bottalico2016air}. Among regulated pollutants, nitrogen dioxide (NO$_2$) is considered to be one of the most hazardous. Short-term exposure to high concentrations is associated with lung inflammation and airway irritation, while long-term exposure has been linked to chronic respiratory illnesses \citep{orellano2020short}. The serious consequences of the pollutant motivate the improvement of methods for characterizing the spatiotemporal distribution of its concentration. 

Networks of ground-based monitoring stations provide high-quality, point-level measurements of air pollutant concentrations and are widely regarded as reliable observational data. However, these networks are inherently biased and irregular in space, with station locations often concentrated in areas with high population density. Coverage is often limited in rural regions and complex terrain, and the set of active stations varies over time due to maintenance or operational constraints. Numerical chemical transport models (CTMs), such as the Community Multiscale Air Quality (CMAQ) model developed by the U.S. Environmental Protection Agency \citep{CMAQ}, and reanalysis products that combine ensembles of CTMs with observational data through data assimilation, such as the European air quality reanalysis produced by the \citet{CAMS} (CAMS), can provide a large-scale view of air pollution dynamics. These models simulate atmospheric transport and chemical processes using emissions inventories and meteorological inputs, producing spatially and temporally complete pollutant concentrations on regular grids. Although CTM outputs offer uniform spatial coverage and capture large-scale spatiotemporal dynamics, their relatively coarse resolution can lead to bias or over-smoothing at local scales.

In this work, we develop a scalable, transfer-learning framework that allows us to integrate the gridded CTM outputs with irregular, point-level monitoring data. The CTM outputs are used to inform a prior covariance that is subsequently adjusted when fitting the spatial model to point-level data. Our ultimate objective is to produce accurate, daily estimates of $\text{NO}_2$ concentrations on a fine grid over all of Italy, for the year 2023. Achieving this goal introduces three methodological challenges. First, the spatial domain and temporal frequency imply very large datasets that lead to computational bottlenecks for classical Gaussian process methods. Second, Italy's geography includes coastlines, mountain ranges, and heterogeneous emission regions, motivating a flexible, non-stationary, and anisotropic model. Third, we must explicitly accommodate the change of support between gridded CTM output and point-level stations in a way that preserves interpretability and allows prediction on a target fine grid.

We choose to model the NO$_2$ concentrations using the LatticeKrig framework \citep{Latticekrig, LKrigpackage}, a basis function method that provides an efficient Gaussian process approximation. The framework uses sparse linear algebra in key matrices to enable working with large spatial data, and the use of basis functions allows for interpolation and prediction across different spatial supports. Additionally, the LatticeKrig model allows for highly flexible, nonstationary, and anisotropic covariance function through the direct specification of a sparse precision matrix. Although this flexibility is attractive, it introduces a substantial inference challenge: the need to estimate a large number of spatially varying covariance parameters. As maximum likelihood estimation (MLE) is computationally prohibitive at the scale considered here, we adopt the LatticeVision framework \citep{LatticeVision}, which uses an amortized, neural, image-to-image (I2I) estimation approach. 

The change of support problem is explicitly addressed by two key extensions. First, we extend amortized parameter estimation \citep{zammit2024neural} to the change-of-support setting by incorporating the basis-function representation of the original LatticeKrig formulation into the LatticeVision parameter estimation pipeline. This extension is essential for change-of-support problems, as the basis functions allow for the transfer of information across different spatial resolutions and supports. In our application, we transfer information between CTM fields and point-level station data, as well as predict onto fine grids. Second, we introduce a likelihood-based refinement that adjusts the covariance structure when transitioning between data sources. In our setting, the CTM outputs capture large-scale dependence but tend to under-represent local variability. We therefore treat CTM-derived anisotropy parameters as informative at regional scales and adjust the local correlation range using station data, combining amortized neural estimation for large, regular grids with likelihood-based refinement for smaller, irregular spatial data.

The remainder of the paper is organized as follows. Section~\ref{sec2} reviews related literature on change-of-support modeling, spatial data fusion, scalable Gaussian processes, and neural parameter estimation. Section~\ref{sec3:data} describes the station, CTM, and auxiliary covariate datasets. Section~\ref{sec4} introduces our statistical model and the nonstationary LatticeKrig formulation, along with the neural parameter estimation approach. Section~\ref{sec5} presents CTM-based parameter estimation and associated experiments. Section~\ref{sec:station} applies the proposed method to the station data, including cross-validation and fine-grid prediction. Section~\ref{sec-discussion} concludes with a discussion and directions for future work. All of the supplementary code and data used for the experiments in this paper are publicly available at \href{https://github.com/afustamo/nonstat-amort-latticekrig}{github.com/afustamo/ItalyAQ-amortized-transfer-learning}.

\section{Related Works}
\label{sec2}

The need to integrate spatial data sources that do not share the same spatial support is often referred to as a change-of-support problem (COSP; \citealt{gotway2002combining}). The transition from point-level data to gridded areal data, and vice versa, was initially addressed through the use of block kriging \citep{banerjee2003hierarchical, cressie2015statistics}. Subsequent data fusion approaches interpreted measurements collected over different spatial supports as distinct observations of the same latent process. In the context of air pollution, \cite{fuentes2005model} expanded upon this approach with a Bayesian framework, estimating the true process from station measurements and CTM outputs. More recently, a closely related approach was adopted by \cite{forlani2020joint}, who investigated NO$_2$ concentrations in Italy during the COVID-19 pandemic. However, Bayesian techniques often involve substantial computational costs, mainly due to complex posterior distribution computation. A second line of methodology aimed at integrating information collected over different spatial supports relies on regression-based approaches. Here the response variable is represented by the most reliable measurements (i.e., ground monitoring stations), and covariates consist of additional information sources such as climate variables and CTM outputs. These covariates are often characterized by higher uncertainty but broader spatial coverage. These methods are commonly referred to as calibration approaches \citep[e.g.,][]{gelfand2003spatial,berrocal2010spatio,berrocal2012space,malings2024air}. To account for local discrepancies between ground-based measurements and gridded data, spatially varying regression coefficients have been introduced \citep{gelfand2003spatial,berrocal2010spatio,berrocal2012space}. 

More recently, other approaches have emerged that incorporate machine learning techniques to predict ground-level concentrations from CTM outputs or satellite observations. These include tree-based methods \citep{stafoggia2019estimation, shetty2024estimating,sun2024inferring} and neural networks \citep{ghahremanloo2021deep}. Although satellite observations have been shown to improve performance in some settings, their overpass times for Italy typically occur during periods of low traffic and correspond to daily minima of NO$_2$ concentrations, making them less representative of daily average pollution patterns. Our approach is most closely related to the transfer-learning framework of \citet{gong2025nonstationary}, in which gridded CTM outputs inform a prior covariance that is subsequently adjusted when fitting the model to point-level data. 

Beyond change-of-support considerations, the spatial domain and temporal frequency in modern environmental problems often comprise large datasets that lead to computational bottlenecks for classical Gaussian process methods \citep{stein2008modeling, sungeostat}. A substantial literature addresses scalability through approximations and sparse representations, including Vecchia-type factorizations \citep{vecchia1988estimation,datta2016hierarchical}, covariance tapering \citep{furrer2006covariance, bevilacqua2016covariance}, spectral representations \citep{fuentes2007approximate}, low-rank methods such as fixed-rank kriging \citep{cressie2008fixed}, SPDE/GMRF formulations \citep{lindgren2011explicit,rodeschini2026multivariate}, and lattice-based models \citep{Latticekrig}. \citet{heaton2019case} assessed many of these techniques in a competition, where the application of the models to the same case study enabled a fair comparison among them.

Big environmental data often exhibit nonstationarity due to heterogeneous terrain and complex meteorological drivers. Modeling this structure requires estimating many spatially varying parameters, rendering classical maximum likelihood estimation computationally infeasible at scale. Neural parameter estimation has recently emerged as a powerful alternative to MLE, where the cost of training the neural network is amortized by its repeated use \citep{liu2020task,zammit2024neural}. Initial approaches for large domains relied on local neural estimators, which used simple dense or convolutional networks to infer parameters at each location or subsection of the domain independently, effectively assuming local stationarity \citep[e.g.,][]{banesh2021fast,gerber2021fast,lenzi2023neural,sainsbury2024likelihood,rai2024fast,rai2025extmodeling}. Recent work has leveraged the observation that in regularly gridded settings, both ensembles of spatial fields and their corresponding spatially varying parameter ``fields'' can be represented as images with multiple channels \citep{LatticeVision,walchessen2025neural}. The LatticeVision framework uses this representation to move beyond local estimation and instead use modern I2I (image-to-image) architectures such as vision transformers to estimate thousands of spatially varying parameters across the entire domain in a single forward pass. Our work builds most directly on this line of research, extending it to explicitly handle change-of-support settings.

%%%%%%%%%%%%%%%%%%% DATA %%%%%%%%%%%%%%%%%%%%%
\section{Data}
\label{sec3:data}

\subsection{Air quality data}
\label{sec:data_aq}
We use two complementary sources of NO$_2$ concentration data: observations from air pollution monitoring stations and outputs from chemical transport models. The station data are obtained from the AQCLIM dataset \citep{fusta2026}, an open-access resource which contains daily measurements of air pollutant concentrations across Italy in the years 2013--2023. The AQCLIM dataset is derived from observations provided by the European Environmental Agency (EEA) air quality monitoring network and processed into a cleaned, harmonized format suitable for statistical analysis. In this study, we focus on daily mean NO$_2$ concentrations for the year 2023. An example daily realization (April 13, 2023) is shown in Figure~\ref{fig:placeholder}. It is important to note that the monitoring network is not fixed over time, as the set of operating stations may vary slightly across days.

The second source of NO$_2$ concentrations is derived from an ensemble of data assimilation systems that rely on Chemical Transport Models (CTMs). CTMs combine emission data with external factors such as wind, temperature, and humidity to simulate the behaviour of air pollutants in the atmosphere. Observations from satellites and monitoring networks are also assimilated into these systems as additional information. We refer to this source as ``CTM data'' throughout the remainder of the paper. The CTM data are sourced from the \citet{CAMS} (CAMS), providing hourly average concentrations of various pollutants on a complete, uniform $0.1^{\circ}\times0.1^{\circ}$ grid across the European continent. For this study, we compute daily mean NO$_2$ concentrations for the year 2023 over the spatial domain covering Italy, spanning longitudes $6.1^{\circ}-18.8^{\circ} \text{E}$ and latitudes $35.2^{\circ}-47.9^{\circ} \text{N}$. An example concentration field for April 13, 2023 is shown in Figure \ref{fig:placeholder}. In addition to NO$_2$ concentrations, we also use a variety of covariates which are described in the following section.

\subsection{Additional covariate data}
\label{sec:addcov}

\begin{figure}[htpb]
    \centering
    \includegraphics[width=0.99\linewidth]{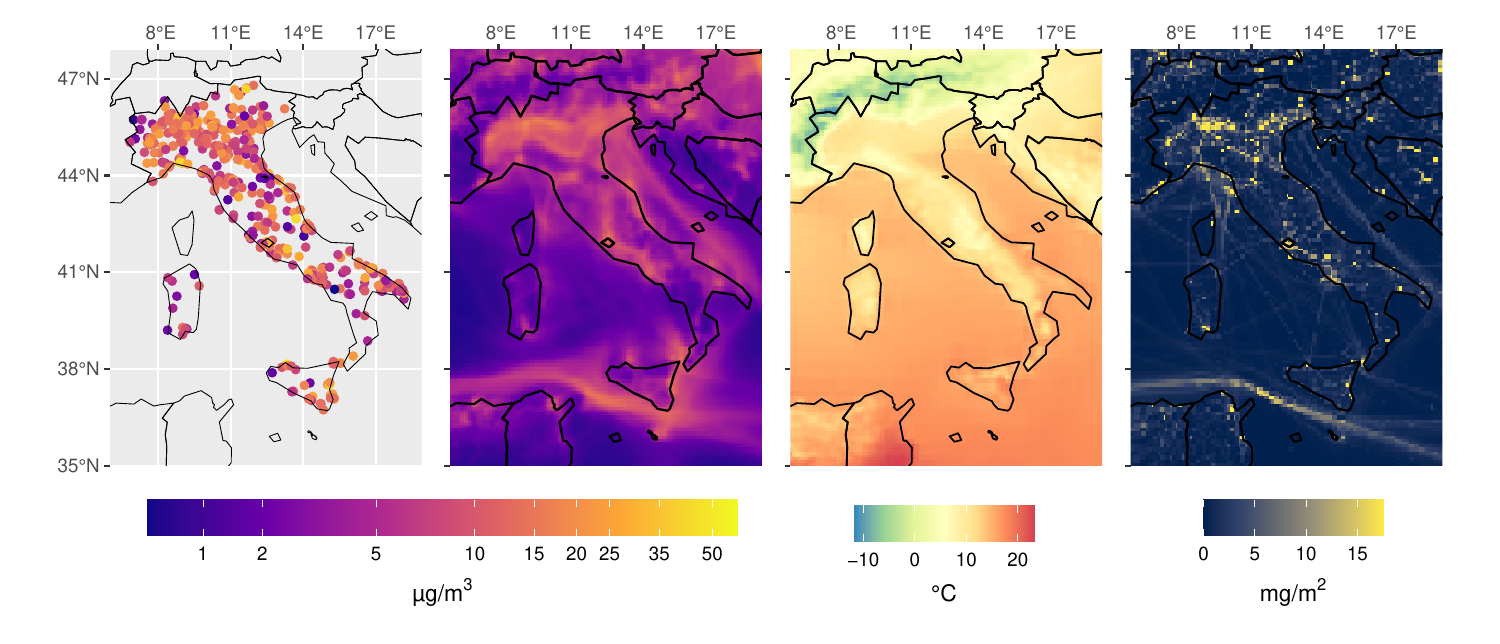}
    \caption{From the left: NO$_2$ concentrations from ground monitoring stations, NO$_2$ concentrations from chemical transport models, temperature from reanalysis (ERA5 datasets), NO$_x$ emission fluxes from CAMS-REG-ANT. Data for Italy, on 13th April 2023.}
    \label{fig:placeholder}
\end{figure}

The chosen meteorological variables represent external drivers of atmospheric transport and dispersion, and are commonly used in air-quality modeling studies \citep[e.g.,][]{ignaccolo2014kriging,otto2024spatiotemporal}. We obtain these covariates from the ERA5 family of reanalysis datasets \citep{hersbach2018era5}, which are generated by the European Centre for Medium-Range Weather Forecasts (ECMWF). We use ERA5-Land \citep{munoz2021era5} as our primary data source, which provides hourly land-surface variables at a spatial resolution of $0.1^{\circ}\times0.1^{\circ}$ and is produced by downscaling the original ERA5 fields. In regions not covered by ERA5-Land, such as coastal areas and small islands, we fill the gaps with ERA5, which provides hourly atmospheric and land-surface variables at a $0.25^{\circ}\times0.25^{\circ}$ resolution. From these data products, we find the daily averages for temperature, relative humidity, surface solar radiation, wind speed, and boundary layer height. An example temperature field for April 13, 2023 is shown in Figure~\ref{fig:placeholder}.

% \subsection{Emission data}
Air pollutant concentrations are largely driven by emissions from ground-based sources. To capture such information, we use the CAMS-REG-ANT dataset \citep{Kuenen2022_CAMSREG}, a gridded anthropogenic emission inventory for Europe based on national emission reports. National totals are spatially distributed onto a 0.05$^{\circ}\!\times$0.1$^{\circ}$ grid using auxiliary information such as facility locations and land-surface typology. We combine the annual gridded emissions with detailed temporal profiles from the CAMS-REG-TEMPO dataset \citep{Guevara2021_CAMSREG}, which provides sector-specific temporal profiles at monthly, weekly, daily, and hourly scales. Combining these two datasets enables the extension of annual totals to daily estimates. We select daily values of NO\textsubscript{x} emissions (NO $+$ NO$_2$) as a key covariate, as primary emissions from ground-based sources are dominated by NO, which is converted to NO$_2$ in the atmosphere due to the presence of volatile organic compounds. An example emission field for April 13, 2023 is shown in Figure~\ref{fig:placeholder}.

% \subsection{Topographic data}
Atmospheric NO$_2$ concentration is closely related to elevation. Low elevation, enclosed regions such as the Po Valley are prone to pollutant accumulation, while high elevations tend to exhibit lower concentrations. We obtain an elevation covariate from the Copernicus Digital Elevation Model (DEM), originally available at 10m$^2$, and aggregate it by averaging to the same spatial resolution of the final prediction grid (i.e., 0.05$^{\circ}\!\times$0.05$^{\circ}$). Together, these seven covariates characterize known, large-scale drivers of NO$_2$ concentrations, and will be incorporated into the linear mean component of our spatial modeling scheme. 

    %PLOT OF ALL INPUT DATA FOR ONE DAY

%%%%%%%%%%%%%%%%%%% METHODS %%%%%%%%%%%%%%%%%%%%%
\section{Methods}
\label{sec4}

In this section, we develop a statistical framework for data fusion in which the objective is to model an underlying spatio-temporal latent process. We observe this process through two distinct data sources: sparse, irregular point-level measurements from monitoring stations, and spatially complete numerical model outputs defined on a coarse regular grid. Both are treated as noisy observations of the same latent field, but differ in spatial support and measurement error structure. Our approach follows a transfer-learning strategy: first, we learn the large-scale, nonstationary, and anisotropic spatial correlation structure from the gridded data using an image-to-image (I2I), neural estimator. We then transfer this structure to the geostatistical model when fitting to the point-level observations, allowing for an additional refinement to account for the change of support problem.

\subsection{The Latent and Observed Processes}

Consider a real-valued spatio-temporal latent process defined over a continuous spatial domain and discrete time (i.e., $\{y_t(\mathbf{s}) : \mathbf{s} \in \mathcal{S} \subset \mathbb{R}^2, t \in \mathcal{T} \subset \mathbb{N}^+ \}$). For a spatial location $\mathbf{s} \in \mathcal{S}$ and day $t \in \mathcal{T}$, $y_t(\mathbf{s})$ follows an autoregressive model where  

\begin{equation}
    y_t(\mathbf{s}) = \mu_t(\mathbf{s}) + \alpha_t y_{t-1}(\mathbf{s}) + g_t(\mathbf{s}),
    \label{eq:arx1_y}
\end{equation}
where $\mu_t(\mathbf{s})$ represents the exogenous fixed effects, $\alpha_t y_{t-1}(\mathbf{s})$ captures the temporal dynamics, and $g_t(\mathbf{s})$ is a mean zero spatial process. The exogenous fixed effects are modeled as $\mu_t(\mathbf{s}) = \mathbf{x}_t(\mathbf{s})^{\top} \boldsymbol{\beta}_t$, with covariates $\mathbf{x}_t(\mathbf{s})$ and regression coefficients $\boldsymbol{\beta}_t$. The residual spatial field $g_t(\mathbf{s})$ is modeled as a Gaussian process $g_t(\mathbf{s}) \sim \mathcal{GP}\!\left(0,\, C_{t}(\mathbf{s},\mathbf{s}')\right)$ where the covariance function, $C_{t}(\mathbf{s},\mathbf{s}')$, is allowed to be nonstationary and anisotropic. We observe two distinct processes, $z_t^{(1)}(\mathbf{s})$  and $z_{i,t}^{(2)}$, that provide noisy measurements of the latent process, each observed on a different spatial support and subject to its own measurement error.

The process $z_t^{(1)}(\mathbf{s})$ is a sparse and irregular point process that is observed at $n_t$ spatial locations that can vary with $t$.
% (i.e., the locations of the monitoring stations). 
We let
\begin{equation}
    z_t^{(1)}(\mathbf{s}) \equiv y_t(\mathbf{s}) + \varepsilon^{(1)}_t(\mathbf{s})
\end{equation}
where $\varepsilon^{(1)}_t(\mathbf{s}) \sim \mathcal{N}(0, \tau_t^2)$ is an independent measurement error term. In complement, $z_{i,t}^{(2)}$ is regularly gridded and has no missing values. The process $z_{i,t}^{(2)}$ is spatially defined on a discretised domain $\mathcal{S}^d$ made of non-overlapping areal units $A_i$, and $\mathcal{S}= \bigcup_{i=1}^{N} A_i$, (i.e., $\mathcal{S}^d \equiv \{ A_i \subset \mathcal{S}:i=1,\ldots,N\}$). The process $z_{i,t}^{(2)}$ represents areal data and is defined as the areal averages
\begin{equation}
    z_{i,t}^{(2)} \equiv \frac{1}{|A_i|} \int_{A_i} y_t(\mathbf{s})\,\text{d}\mathbf{s} + \varepsilon^{(2)}_{i,t}
\end{equation}
where $\varepsilon^{(2)}_{i,t} \sim \mathcal{N}(0, \xi_t^2)$ is an error term.

\subsubsection{Data Fusion and Change of Support Problem}
\label{sec:datafusion_cosp}
Data fusion and change of support problems (COSP) arise from the relationship between point-level data $z_{t}^{(1)}(\mathbf{s})$ and areal data $z_{i,t}^{(2)}$. Given the full spatial and temporal coverage of the areal data, we aim to transfer information from $z_{i,t}^{(2)}$ to the point-level model, thus improving our understanding of the hidden process. Specifically, we use areal data to learn the spatial correlation structure and transfer it to the point-level model as a prior estimate of the covariance function. The connection between $z_{t}^{(1)}(\mathbf{s})$ and $z_{i,t}^{(2)}$ is detailed in \ref{app:cosp} and a brief description is provided here. The model for the areal data can be written as
\begin{equation}
    z_{i,t}^{(2)} = \bar{\mathbf{x}}_{i,t}^{\top}\boldsymbol{\beta}_t+ \alpha_t z_{i,t-1}^{(2)} + \eta^{(2)}_{i,t},
    \label{eq:arx1_z2}
\end{equation}
with
\begin{equation}
    \eta^{(2)}_{i,t} = \bar{g}_{i,t} + \varepsilon_{i,t}^{(2)} + \alpha_t\varepsilon_{i,t-1}^{(2)},
\end{equation}
where $\bar{\mathbf{x}}_{i,t} = |A_i|^{-1} \int_{A_i} \mathbf{x}_t(\mathbf{s})\,\text{d}\mathbf{s}$ and $\bar{g}_{i,t} = |A_i|^{-1} \int_{A_i} g_t(\mathbf{s})\,\text{d}\mathbf{s}$.
Therefore, $\eta^{(2)}_{i,t}$ captures the spatial component and the error term, in particular the collection $\bm{\eta}^{(2)}_{t} = (\eta^{(2)}_{1,t},\ldots,\eta^{(2)}_{N,t})^{\top}  \sim \mathcal{N}(0,\Sigma_{\eta_t}^{(2)})$ with $\Sigma_{\eta_t}^{(2)} = \Sigma_{\bar{g}_t}+I(\xi^2_t + \alpha_t^2\xi^2_{t-1}))$. Elements of the matrix $\Sigma_{\bar{g}_t}$ correspond to the covariance between all pairs $\bar{g}_{i,t}$ and $\bar{g}_{j,t}$. From $\bm{\eta}^{(2)}_t$ parameters of $\Sigma_{\bar{g}_t}$ are found using a neural estimator, discussed in Section~\ref{subsec:neural_param_est}. 

To establish the connection between areal and point-level data, we decompose $y_t(\mathbf{s})$ into the areal average $\bar{y}_{i,t}$ and a residual term $\psi_t(\mathbf{s})$ that captures the fine-scale variability that is smoothed out by areal averaging. Specifically, for any $\mathbf{s} \in A_i$,
\begin{equation}
    y_t(\mathbf{s})=\bar{y}_{i,t} + \psi_t(\mathbf{s}),
\end{equation}
where $\psi_t(\mathbf{s}) \sim GP(0,C_{\psi_t}(\mathbf{s},\mathbf{s'}))$ and $C_{\psi_t}$ is a covariance function such that $C_{\psi_t} = f(C_{g_t})$. The point-level model can be rewritten as
\begin{equation}
    z_t^{(1)}(\mathbf{s}) = \bar{\mathbf{x}}_{i,t}^{\top}\boldsymbol{\beta}_t + \alpha_t z_{i,t-1}^{(2)} + \eta^{(1)}_{t}(\mathbf{s}),
\end{equation}
with
\begin{equation}
    \eta^{(1)}_{t}(\mathbf{s}) = \bar{g}_{i,t} + \varepsilon^{(1)}_t(\mathbf{s}) - \alpha_t\varepsilon^{(2)}_{i,t-1} + \psi_t(\mathbf{s}),
\end{equation}
where $\eta^{(1)}_{t}(\mathbf{s}) \sim GP(0,C_{\eta_t}^{(1)})$. The form of $C_{\eta_t}^{(1)}$ is described in \ref{app:cosp}. 
In summary, both the point-level covariance function $C_{\eta_t}^{(1)}(\cdot,\cdot)$ and the areal covariance matrix $\Sigma_{\eta_t}^{(2)}$ encode the same large-scale spatial structure determined by $\Sigma_{\bar{g}_t}$, justifying a transfer learning approach between the two data sources. Moreover, for a sufficiently small $|A_i|$, the resulting covariances are approximately equivalent. In the case of larger $|A_i|$, $\Sigma_{\eta_t}^{(2)}$ does not include the short-range structure from $C_{\psi_t}$. As it is unclear whether $|A_i|$ is sufficiently small in our application, we account for this effect by estimating an additional parameter in the point-level model, as detailed in Section \ref{sec:cosp_met}

\subsection{The Non-stationary LatticeKrig Model}
\label{sec:nonstat_LK}

The LatticeKrig model specifies a GP through an efficient basis-function representation with coefficients governed by a spatial autoregression (SAR). These ingredients result in spatial dependence being controlled by a sparse precision matrix and provide scalable computations for large spatial data. By the LatticeKrig specification, the spatial component  $g_t(\mathbf{s})$ becomes $\Phi(\mathbf{s})^\top\mathbf{c}_t$ and the model for the true latent process for a single day becomes:

\begin{equation}
    y_t(\mathbf{s}) 
        = \mathbf{x}_t(\mathbf{s})^\top \boldsymbol{\beta}_t
        + \alpha_t y_{t-1}(\mathbf{s})
           + \Phi(\mathbf{s})^\top\mathbf{c}_t
           ,
\end{equation}
where $\Phi(\mathbf{s})^\top \equiv (\phi_1(\mathbf{s}),\ldots,\phi_m(\mathbf{s}))$ are the set of basis functions evaluated at location $\mathbf{s}$ and $\mathbf{c}_t \equiv (c_{1,t},\ldots,c_{m,t})^\top$ are the coefficients. For ease of exposition, we consider a single day, omitting $t$ from our notation.

The basis functions \(\phi_j(\mathbf{s})\) are compactly supported, radially symmetric Wendland functions centered at $m$ lattice locations $\{\mathbf{u}_j\}_{j=1}^m \in \mathcal{S}$, and the coefficients satisfy \(\mathbf{c} \sim \mathcal{N}(\mathbf{0}, Q^{-1})\), where $Q$ is the sparse precision matrix. The precision matrix $Q$ is specified through a spatial autoregressive (SAR) representation. Given a sparse linear operator $B$ and a random vector $\mathbf{e} \sim \mathcal{N}(\mathbf{0}, I)$, we construct the distribution of $\mathbf{c}$ such that $B\mathbf{c} = \mathbf{e}$. Thus, the operator $B$ is a SAR matrix that decorrelates the coefficients to white noise. In the stationary case, the rows of the matrix $B$ are filled according to the five-point stencil 
\begin{equation}
    \begin{array}{c|c|c}
 & -1 & \\
\hline
-1 & 4+\kappa^2 & -1\\
\hline
 & -1 &
\end{array}.
\end{equation}
The entries of the stencil determine the nonzero coefficients in each row of $B$: the center value corresponds to the diagonal entry, the four off-center values correspond to the immediate neighbors on the lattice, and all remaining entries in the row are zero. Here, $\kappa^2$ controls the correlation range for the spatial domain. By linear statistics, $\mathbf{c} \sim \mathcal{N}(\mathbf{0}, (B^{\top} B)^{-1})$, thus the corresponding precision matrix $Q = B^{\top}B$ is also sparse. 

For the LatticeKrig model, nonstationarity and anisotropy are achieved by expanding the SAR stencil to second order neighbors and allowing the stencil weights to vary across the domain. Specifically, we allow the diagonal and off-diagonal entries of $B$ to depend on interpretable and spatially varying parameters $\theta(\mathbf{u})$, $\rho(\mathbf{u})$ and $\kappa^2(\mathbf{u})$. Unlike prior work by \citet{LatticeVision} and \citet{ wiens2020modeling}, these parameters vary across the  lattice of basis functions centers $\{\mathbf{u}_j\}_{j=1}^m$, rather than the spatial locations $\{\mathbf{s}_i\}_{i=1}^N$.
At each basis-function center $\mathbf{u}_j$, the parameter $\kappa^2(\mathbf{u}_j)$ controls the correlation range (i.e., a larger $\kappa^2(\mathbf{u}_j)$ means smaller correlation range), the parameter $\rho(\mathbf{u}_j)$ controls the degree of anisotropy (e.g., $\rho(\mathbf{u}_j) = 1$ corresponds to isotropy at $\mathbf{u}_j$), and $\theta(\mathbf{u}_j)$ controls the direction of anisotropy. The anisotropy parameters are used to build the rotation matrix $\Psi(\mathbf{u})$ and the anisotropic scaling matrix $\Lambda(\mathbf{u})$,

\begin{equation}
\Psi(\mathbf{u}) =
\begin{bmatrix}
\cos\theta(\mathbf{u}) & -\sin\theta(\mathbf{u})\\
\sin\theta(\mathbf{u}) &  \cos\theta(\mathbf{u})
\end{bmatrix},\quad
\Lambda(\mathbf{u}) =
\begin{bmatrix}
\sqrt{\rho(\mathbf{u})} & 0\\
0 & \frac{1}{\sqrt{\rho(\mathbf{u})}}
\end{bmatrix},
\label{eq:rot_diag_mat}
\end{equation}
which are then used to create the $2\times2$ dispersion matrix: $D(\mathbf{u}) = \Psi(\mathbf{u})^{\top}\Lambda(\mathbf{u})\Psi(\mathbf{u})$. The entries of $B$ in the nonstationary case are filled according to the new spatially varying nine-point stencil
\begin{equation}
\begin{array}{c|c|c}
\frac{D_{1,2}(\mathbf{u})}{2} & -D_{2,2}(\mathbf{u}) & \frac{-D_{1,2}(\mathbf{u})}{2} \\
\hline
-D_{1,1}(\mathbf{u}) & \kappa^2(\mathbf{u}) + 2 D_{1,1}(\mathbf{u}) + 2 D_{2,2}(\mathbf{u}) & -D_{1,1}(\mathbf{u}) \\
\hline
\frac{-D_{1,2}(\mathbf{u})}{2} & -D_{2,2}(\mathbf{u}) & \frac{D_{1,2}(\mathbf{u})}{2}
\end{array} .
\label{eq:anisotropic_stencil}
\end{equation}
This formulation retains computational advantages, specifies a rich class of non-stationary and anisotropic GPs, and avoids the difficult specification of an analytical covariance function. Another key advantage is that the spatially varying parameters are interpretable and can yield physical insights into the spatial dependence of the field. At each basis-function center, $\mathbf{u}_j$, we can represent the combination of the parameters as an elliptical dependence contour, where the orientation is
governed by $\theta(\mathbf{u}_j)$, the aspect ratio is controlled by $\rho(\mathbf{u}_j)$, and the overall size is controlled by $\kappa^2(\mathbf{u}_j)$. It is important to note that this specification is with respect to the SAR, and when the implied covariance is evaluated, the interplay between these elliptical profiles can result in very complex and flexible spatial dependence structures. 

However, the estimation of the resulting, spatially varying parameter fields $(\boldsymbol{\kappa}^2, \boldsymbol{\rho}, \boldsymbol{\theta})$ is a challenging inference problem, especially for large datasets. Fitting the model directly via maximum likelihood or Bayesian methods is not feasible. To address this problem, we use the amortized, neural parameter estimation strategy from the LatticeVision framework \citep{LatticeVision}. Here, a neural network learns a global mapping from ensembles of spatial fields to the associated parameter fields. This approach enables fast and accurate one-shot estimation of the parameters in seconds and is described in the following sections. 

\subsection{Neural Parameter Estimation}
\label{subsec:neural_param_est}

We use the neural, image-to-image (I2I) estimation strategy from the LatticeVision framework. Specifically, we use the STUN estimator, which is a hybrid architecture that combines the U-Net and Vision Transformer architectures to leverage both the convolution and attention mechanisms. The key feature underlying this approach is that both the input (spatial fields) and the output (parameter fields) are naturally arranged on a regular grid and can therefore be treated as images \citep{LatticeVision}. This method has been shown to outperform local neural estimators with simpler architectures in both computational efficiency and accuracy. Additionally, these I2I architectures are amortized. Training them takes significant computational resources, but this cost is recouped upon deployment, as the trained net maps an ensemble of spatial fields directly to spatially varying parameter fields in a single forward pass.

\subsubsection{Training Procedure}
We follow the synthetic data generation and training procedure of \cite{LatticeVision} with a few key modifications. In brief, we sample from scientifically motivated spatial pattern priors to generate three parameter fields
\[
P = \big[\boldsymbol\kappa^2, \boldsymbol\rho, \boldsymbol\theta\big]^{\top},
\]
on the basis function lattice $\{\mathbf{u}\}_{j=1}^m$. We stack the parameter fields along the channel dimension, treating each field as an image, yielding a target tensor $P \in \mathbb{R}^{3 \times h_\mathbf{u} \times w_\mathbf{u}}$, where $h_\mathbf{u} \times w_\mathbf{u} = m$. These parameter fields define a nonstationary SAR matrix \(B\), as described in Section \ref{sec:nonstat_LK}. To create training inputs, we draw independent white-noise vectors \(\mathbf{e}^{(k)} \sim \mathcal{N}(\mathbf{0}, I)\) and solve the linear systems
\[
B\,\mathbf{c}^{(k)} = \mathbf{e}^{(k)}, \qquad k = 1,\ldots,r,
\]
yielding coefficient vectors \(\mathbf{c}^{(k)} \sim \mathcal{N}(\mathbf{0}, Q^{-1})\) with precision matrix \(Q = B^\top B\). Each coefficient vector is then mapped to a spatial field via the LatticeKrig basis expansion
\[
g^{(k)}(\mathbf{s}) = (\Phi \mathbf{c}^{(k)})(\mathbf{s}),
\]
producing \(r\) independent realizations that share the same underlying nonstationary and anisotropic covariance structure. Following \cite{LatticeVision}, we standardize these realizations: for each pixel, we subtract its mean and divide by its standard deviation across the $r$ replicates, resulting in each pixel having zero mean and unit variance. Stacking the \(r\) simulated fields yields the input tensor $G = \begin{bmatrix} g^{(1)} & \cdots & g^{(r)}\end{bmatrix} \in \mathbb{R}^{r \times h_A \times w_A}$, where $h_A \times w_A = N$. Given an input $G$, our STUN estimator is trained to minimize a pixelwise regression loss between the predicted ($\widehat{P}$) and true ($P$) parameter images. 

\subsubsection{Implementation and Evaluation}
\label{subsubsec:neural_implement}

For simplicity, we align the basis function lattice and the  data grid, so both the input spatial fields and output parameter fields have size $h_A \times w_A = N = h_\mathbf{u} \times w_\mathbf{u} = m = 128\times128$. When creating the target parameter images $P$, the parameter values are sampled from the following range: $\rho \in [1,7]$, $\theta \in [-\frac{\pi}{2},\frac{\pi}{2})$, and $\text{log}(\kappa^2) \in [-9.2,2.3]$ (with $\kappa^2 \in [10^{-4},10]$ represented on a log scale during training and evaluation for numerical stability). These parameter values were chosen to capture a broad range of spatial relationships, ranging from isotropy ($\rho$ = 1) to very elongated dependence ellipses ($\rho \gg 1$) in any direction. The upper bound of $\kappa^2$ is set to 10 to include shorter correlation ranges in the data. We choose an ensemble size of $r = 30$ replicates within each input $G$, and following the procedure in \citet{LatticeVision}, we generate an entirely synthetic dataset with $20,000$ training pairs $(G,P)$. We use the 90/8/2 (train/validation/test) split, resulting in a test set of 400 synthetic data samples, which we use to verify the accuracy of our estimator. All remaining implementation details are provided in \ref{sec:appendix_STUN}.

Table~\ref{tab:stun_metrics} summarizes validation performance of the neural estimator across all pixels with several standard image regression metrics. We find that STUN recovers all parameters with high accuracy, supporting its use as an efficient surrogate for exact, likelihood-based inference.

\begin{table}[htbp]
\centering
\caption{STUN parameter-estimation performance (all true vs.\ predicted pixels) on held-out test data. Here, NRMSE is computed relative to the range (max - min) of parameter values. Relative error is summarized by the median and is reported as a percentage. Arrows indicate the desirable direction.}
\label{tab:stun_metrics}
\begin{tabular}{lccccc}
\toprule
Parameter & RMSE $\downarrow$ & NRMSE $\downarrow$ & MAE $\downarrow$ & $R^2$ $\uparrow$ & Median Rel.\ Err.\ $\downarrow$ \\
\midrule
$\text{log}(\kappa^2)$ & 0.2169 & 0.01884 & 0.1317 & 0.9898 & 6.87\% \\
$\theta$   & 0.0896 & 0.02987 & 0.0509 & 0.9832 & 6.89\% \\
$\rho$     & 0.3912 & 0.06521 & 0.2779 & 0.9227 & 5.17\% \\
\bottomrule
\end{tabular}
\end{table}

\subsection{Extensions for Data Fusion and Change of Support Problem}
\label{sec:cosp_met}

To address the change-of-support problem (COSP) arising from the integration of areal and point-referenced data, this work introduces two key extensions to the amortized neural estimation framework in \citet{LatticeVision}. First, we make use of the compactly supported basis functions in the original LatticeKrig formulation, with the SAR being applied to the coefficients. The basis functions enable the transfer from gridded to point-level data, along with interpolation to finer grids. Second, we allow for the refinement of the parameter $\kappa^2$, which controls correlation range, when transferring information from the regularly gridded data to the point-level observations, as discussed below. 

The initial estimates of the parameter fields for a particular day $\widehat{P} = \big[\widehat{\boldsymbol{\kappa}}^2, \widehat{\boldsymbol{\rho}},\widehat{\boldsymbol{\theta}}]^{\top}$  are first obtained from the regularly gridded data using STUN. To adjust for the change of support, we assume that the correlation range should decrease because the coarse, gridded data is likely to over-smooth the fine-scale structure that is present in the point-level data (see Section ~\ref{sec:datafusion_cosp} and ~\ref{app:cosp}).

Therefore, we supplement the initial estimate of $\boldsymbol{\kappa}^2$ with an additional, stationary term $\kappa^2_{point}$, which is estimated from the point-level data. The adjusted estimate of $\boldsymbol{\kappa}^2$ is then
\begin{equation}
\widehat{\boldsymbol{\kappa}}^2_\mathrm{adj} = \widehat{\boldsymbol{\kappa}}^2 + \boldsymbol{w} \widehat{\kappa}^2_{point}
\end{equation}
where the weights $\boldsymbol{w} \in \mathbb{R}^m$ can be chosen to allow heterogeneous adjustments. In this paper, point-level data are only present on land. Accordingly, we set $w(\mathbf{u})=1$ for basis-function centers located on land and $w(\mathbf{u})=0$ for centers located over the ocean, where no point-level observations are available. Since the point level data is available at a small number of observations $n_t$, the additional $\kappa^2_{point}$ parameter is estimated via MLE. Regarding anisotropy, we assume the initial estimates are sufficiently informative, thus we do not adjust $\widehat{\boldsymbol{\rho}} \text{ and } \widehat{\boldsymbol{\theta}}$.

%%%%%%%%%%%%%%%%%%% CTM APPLICATIONS %%%%%%%%%%%%%%%%%%%%%

\section{Application to Numerical Model Outputs}
\label{sec5}

\subsection{Parameter Estimation}
\label{sec5_param_est}

The CTM outputs are formatted to a regular \(0.1^{\circ}\times0.1^{\circ}\) grid over longitudes \(6.1^{\circ}\)–\(18.8^{\circ}\)E and latitudes \(35.2^{\circ}\)–\(47.9^{\circ}\)N. Here, \(\{A_i\}_{i=1}^{N}\) denotes the collection of non-overlapping grid cells and \(N=128\times128\). For each day in 2023, \(t \in \{1,\ldots,365\}\), the CTM field is identified as the areal process \(\{z^{(2)}_{i,t}\}_{i=1}^m\), where each \(z^{(2)}_{i,t}\) corresponds to the grid-cell value over \(A_i\). All covariates are also aligned to this grid. When a covariate is available at a different resolution, we upscale to the CTM grid resolution by averaging (e.g., elevation and emissions) or downscale via nearest-neighbor resampling (e.g., ERA5 where ERA5-Land is unavailable). To isolate the residual spatial process $\bar{g}_{i,t}$ for parameter estimation, we fit and then subtract the conditional mean model for each day. Specifically, the mean of our process is the ARX(1) regression, described in Eq. \ref{eq:arx1_z2}, composed of $\bar{\mathbf{x}}_{i,t}^{\top}\boldsymbol{\beta}_t + \alpha_tz^{(2)}_{i,t-1}$, where $\bar{\mathbf{x}}_{i,t}$ includes an intercept, latitude, longitude, and the seven covariates described in Section~\ref{sec:addcov}, and $z^{(2)}_{i,t-1}$ is the lagged $\text{NO}_2$ concentration from the numerical model. The choice to include lagged $\text{NO}_2$ values as a covariate is motivated by a well-known, strong AR(1) behavior in air pollutant concentrations \citep[e.g.,][]{calculli2015maximum}. 

We assume the residual spatial process evolves slowly and smoothly over time. This is supported by the structure of our mean model: the AR(1) term removes the bulk of temporal autocorrelation, while covariates such as wind speed and boundary layer height account for much of the daily variation in anisotropic parameters, leaving residuals that are largely driven by slower seasonal patterns. Neighboring residual fields can therefore be treated as pseudo-replicates from a common distribution, allowing for more accurate parameter estimation and stable covariances. We therefore adopt a moving window strategy, including the 15 days before and the 14 days after the day selected, constructing ensembles of $r=30$ residual fields
\[
G_t \in \mathbb{R}^{30 \times 128 \times 128},
\]
which are centered at day $t$. To avoid temporal boundary effects at the beginning and the end of 2023, we extend our time coverage to the last 15 days of 2022 and the first 14 days of 2024. To align with the synthetic data procedure, within each ensemble $G_t$, the 30 residual fields have been standardized pixel-wise. The zero-mean and unit variance fields are then used as the multi-channel input image to STUN. Following Section~\ref{subsubsec:neural_implement}, we align the grid cells and the lattice of basis function centers, so the parameter fields are estimated on a $128\times128$ lattice of basis-function centers. For each day $t$, STUN outputs 
\[
\widehat{P}_t =
[
\widehat{\boldsymbol\kappa}^2_t, 
\widehat{\boldsymbol\rho}_t, 
\widehat{\boldsymbol\theta}_t
]^{\top}
\in \mathbb{R}^{3 \times 128 \times 128}.
\] 
Overall, this amounts to $365 \times 128 \times 128 \times 3 = 17{,}940{,}480$ parameter values that must be estimated. The amortized nature of our estimator means inference is extremely fast: on a single GPU, STUN executes 365 forward passes to estimate all $17{,}940{,}480$ parameter values in just 4 seconds (averaged over 30 trials). Hardware details can be found in the \ref{sec:appendix_STUN}.

\begin{figure}[h]
\centering
\includegraphics[width=1.0\textwidth, trim={0 1cm 0 0}, clip]{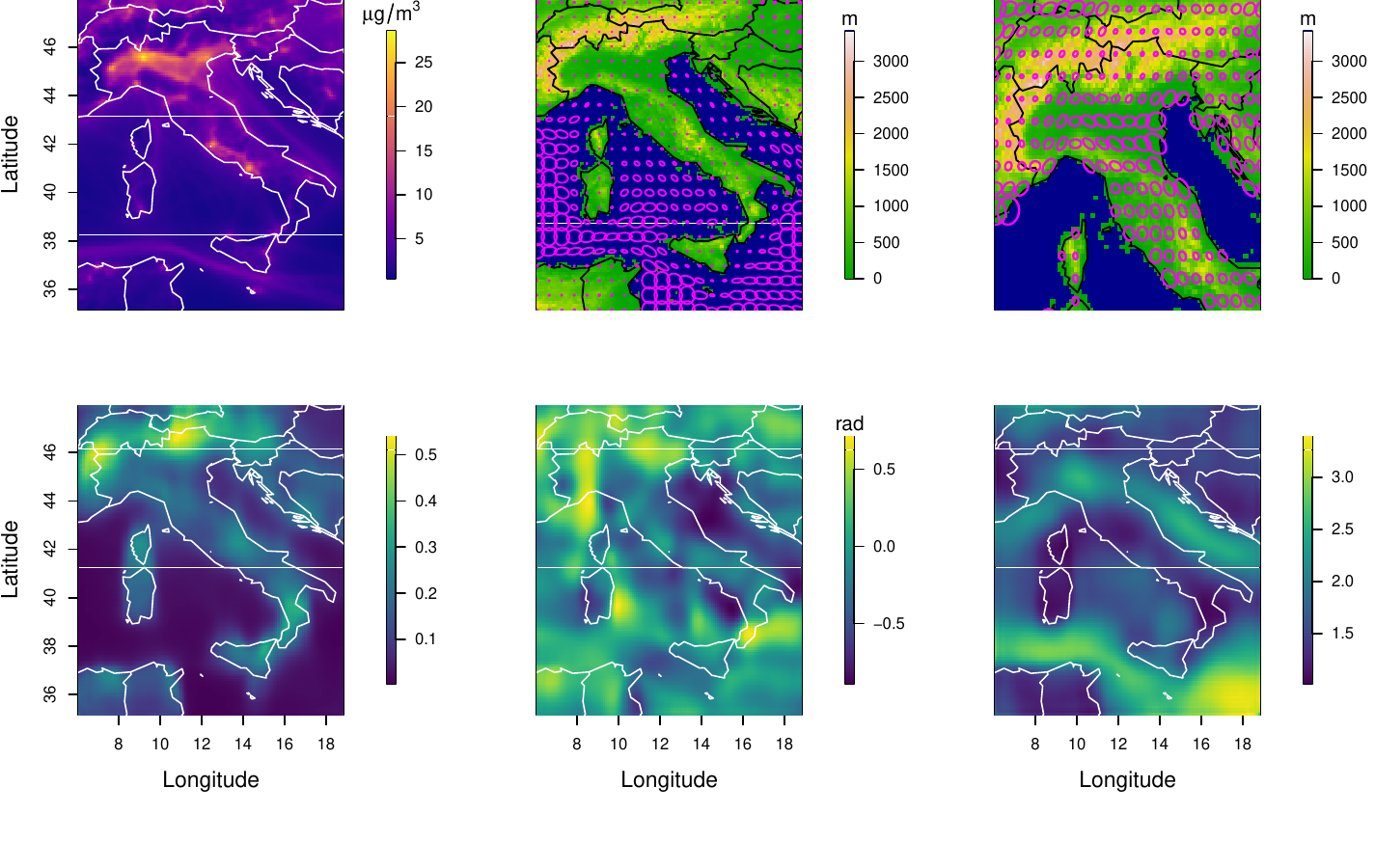}
\caption{2023 annual averages of CTM-based parameter estimates. Top row from left to right: the average CTM NO$_2$ field, elliptical dependence contours overlaid on an elevation (m) map of the full domain, and a zoomed-in view of enlarged ellipses over land only. Bottom row from left to right: annual averages of STUN estimates $\boldsymbol{\widehat{\kappa}}^2$, $\boldsymbol{\widehat{\theta}}$, and $\boldsymbol{\widehat{\rho}}$.}
\label{fig:stun_results}
\end{figure}

Qualitatively, the resulting parameter estimates exhibit physically plausible spatial structure. Figure~\ref{fig:stun_results} summarizes spatial patterns through 2023 annual averages of the $\text{NO}_2$ field, the estimated parameters, and elliptical contours representing the average local dependence structure. The inferred range fields $\{\boldsymbol{\widehat{\kappa}}^2_t\}$ vary smoothly at large scales while preserving sharp transitions near coastlines, with generally shorter ranges over land and longer ranges over surrounding water. The anisotropy fields $\{\boldsymbol{\widehat{\rho}}_t\}$ and $\{\boldsymbol{\widehat{\theta}}_t\}$ highlight directional dependence aligned with major geographic features: we observe pronounced anisotropy and consistent orientation patterns in the Po Valley, along the Adriatic Sea, and following maritime routes. Systematic changes in dependence structure are also evident near complex terrain such as mountain ranges.

\subsection{CTM Reconstruction Experiment}
\label{sec5_ctm_recon}

\begin{figure}[h]
\centering
\includegraphics[width=1.0\textwidth]{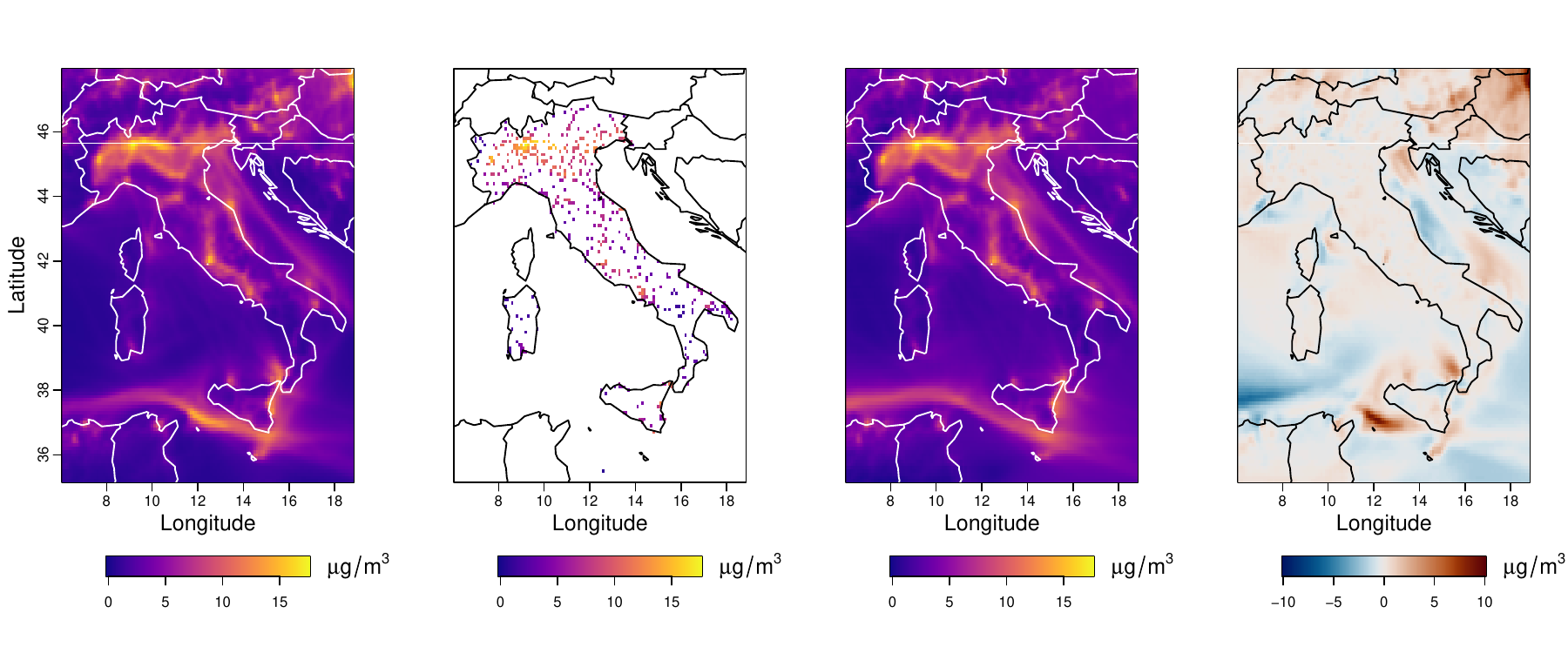}
\caption{CTM NO$_2$ reconstruction experiment for April 13, 2023. From left to right: the actual CTM NO$_2$ field, masked data used for model fitting (pixels at locations with stations), the full prediction from the non-stationary model, and the difference between the true and predicted field.}
\label{fig:ctm_reconstruct_exper}
\end{figure}

Having obtained parameter fields for each day of 2023 from STUN, we assess whether these nonstationary parameters improve predictive performance. To do so, we design a controlled reconstruction experiment using the CTM data. This experiment mirrors our later application to monitoring station data, but with the advantage that, unlike monitoring station data, the full CTM NO$_2$ fields are known. For each day, we mask out all grid cells except those coinciding with locations where monitoring stations exist. On the $128 \times 128$ CTM grid (16,384 pixels), this leaves 483 pixels (approximately 3\%) as observed locations, with the remaining pixels treated as missing and used for evaluation. We then predict the full CTM NO$_2$ field from sparse observations under different covariance assumptions. Figure~\ref{fig:ctm_reconstruct_exper} illustrates the true field and the masked data for a representative day.

We compare two models. First, we fit a stationary, isotropic LatticeKrig model to each day. The mean function of this stationary model is fit using the ARX(1) structure described in Section~\ref{sec5_param_est}. The model uses the same single-resolution basis construction with $m = 128 \times 128$ basis functions as the nonstationary model, and the covariance structure is governed by a single global range parameter $\kappa^2$, estimated by maximum likelihood. This stationary Gaussian process model serves as a baseline, and is representative of standard approaches in spatial statistics. Second, we fit the nonstationary, anisotropic LatticeKrig model, again using the same mean structure and CTM NO$_2$ data, with covariance structure governed by the STUN-estimated spatially varying parameter fields $\boldsymbol{\widehat{\kappa}}^2$, $\boldsymbol{\widehat{\rho}}$, and $\boldsymbol{\widehat{\theta}}$ from Section~\ref{sec5_param_est}.

\begin{figure}[h]
\centering
\includegraphics[width=0.9\textwidth]{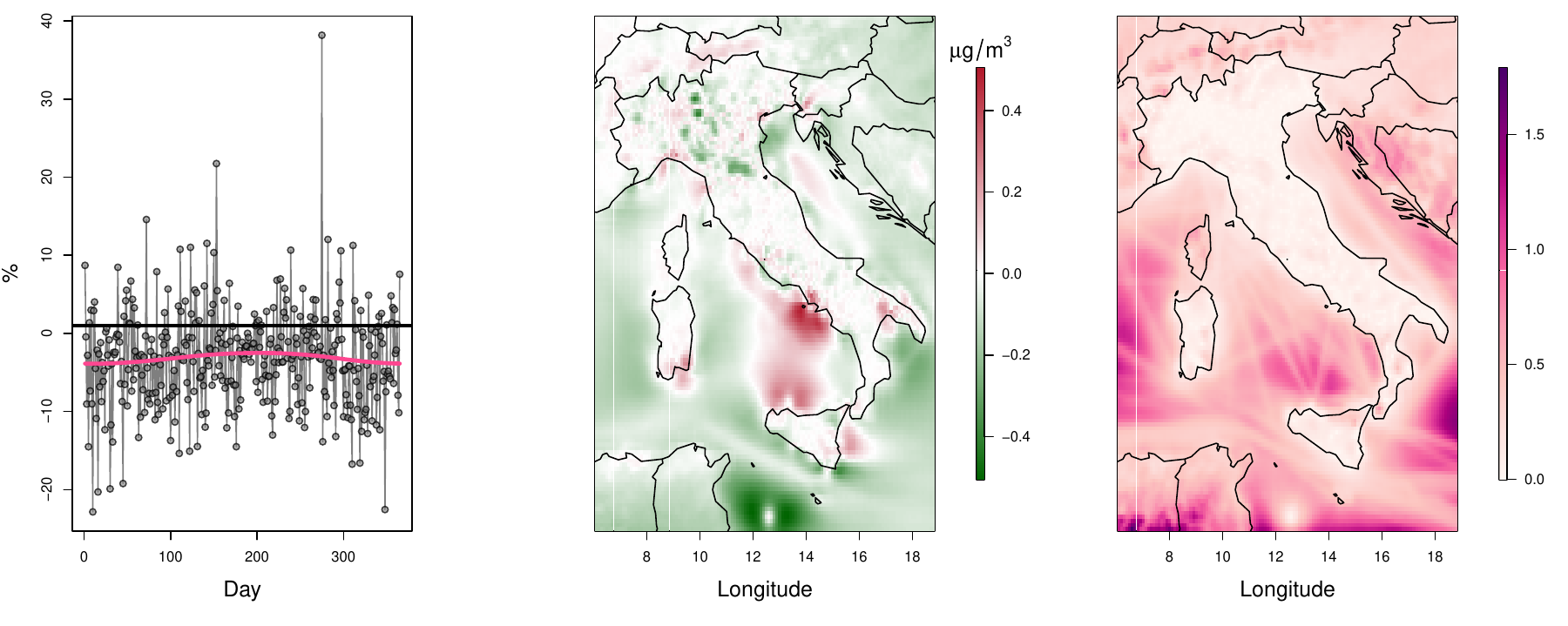}
\caption{ Left: time series of daily RMSE percent differences ((nonstationary / stationary - 1)$\times 100$). A solid black line illustrates equality in both cases, while the pink line is a GAM-based fit with a cyclic cubic spline basis ($20$ basis functions), chosen to capture seasonal structure in the temporal trend. Dots below the black line indicate days on which the nonstationary model outperforms the stationary one. Center: difference in pixel-wise RMSE (nonstationary $-$ stationary); negative values indicate locations where the nonstationary model outperforms the stationary model. Right: the pixel-wise nonstationary RMSE divided by the local annual mean NO$_2$ concentration to show relative errors.}
\label{fig:CTM_ts}
\end{figure}

Across the year, the nonstationary model achieves a lower reconstruction RMSE for the held-out pixels on 267 of 365 days, corresponding to an overall RMSE improvement of 3.2\%. The relative improvement exhibits a seasonal structure, as the nonstationary model tends to outperform the stationary model more strongly during colder months, where daily RMSE improvements can be as large as 22.8\% (see Figure~\ref{fig:CTM_ts}). In absolute terms, the nonstationary model achieves an RMSE of 1.093 $\mu$g/m$^3$ compared to 1.137 $\mu$g/m$^3$ for the stationary model.
% Colder months are associated with a lower and more stable atmospheric boundary layer, which suppresses vertical mixing and allows NO$_2$ to accumulate near the surface. Combined with increased emissions from traffic and residential heating, this leads to higher concentrations and more localized structure that is better captured by the nonstationary model. 
Spatially, the largest relative reconstruction errors tend to occur in mountainous regions outside Italy and along major shipping routes, where emissions are highly variable and Italian monitoring stations are not present. In contrast, errors across Italy remain very small. The difference map in the center panel of Figure~\ref{fig:CTM_ts} shows that the nonstationary model achieves lower RMSE than the stationary model in key regions, including the Po Valley and areas with sparse or nonexistent station coverage. Overall, the results of this experiment provide an argument for the use of STUN-estimated parameters for the subsequent application to monitoring station data.

%%%%%%%%%%%%%%%%%%% STATION APPLICATIONS %%%%%%%%%%%%%%%%%%%%%

\section{Application to Ground-based Measurements}
\label{sec:station}

For 2023, a total of 744 monitoring stations in Italy reported at least one daily NO$_2$ measurement. To ensure sufficient spatial coverage for reliable inference, we restrict our analysis to days with at least $250$ active stations, yielding $349$ valid days. The omitted 16 days exhibited a sharp decline in network coverage, with several days providing only a few dozen observations. Additionally, we utilize only ``background'' stations and remove anomalous measurement values (>80 $\mu g/m^{3}$).

\subsection{Model Estimation}

Similarly to Section \ref{sec5_ctm_recon}, we fit three different models of increasing complexity for each of the $349$ days. The first two models are those used in the reconstruction experiment in Section \ref{sec5_ctm_recon}. The first model is the stationary, isotropic LatticeKrig model with a single global range parameter estimated by maximum likelihood. The second is the nonstationary, anisotropic LatticeKrig model that uses the STUN-derived parameter fields estimated from the CTM data in Section~\ref{sec5_param_est}, but without further correlation range adjustment. The third and most flexible model improves the CTM-informed covariance with an additional range refinement term using the method described in Section \ref{sec:cosp_met}. Specifically, when fitting the model, we define the final $\boldsymbol{\kappa}^2$ field as 
\begin{equation}
\widehat{\boldsymbol{\kappa}}^2_\mathrm{adj} = \widehat{\boldsymbol{\kappa}}_{CTM}^2 + \boldsymbol{w} \widehat{\kappa}^2_{EEA}
\end{equation}
where $\widehat{\boldsymbol{\kappa}}_{CTM}^2$ is the parameter field previously estimated from CTM data using STUN, and $\kappa^2_{EEA}$ is a single parameter that reduces the correlation range and is estimated from the station data by maximum likelihood. The weights $\boldsymbol{w} \in \mathbb{R}^m$ are specified such that $w(\mathbf{u})=1$ for basis-function centers located on land and $w(\mathbf{u})=0$ otherwise. 

\begin{figure}[htbp]
     \centering
     \begin{subfigure}[b]{0.48\textwidth}
         \centering
         \includegraphics[width=\textwidth]{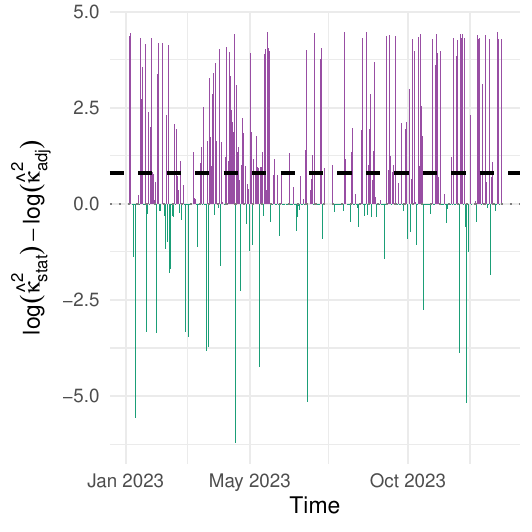}
         \label{fig:kappa2_time}
     \end{subfigure}
     \hfill
     \begin{subfigure}[b]{0.48\textwidth}
         \centering
         \includegraphics[width=\textwidth]{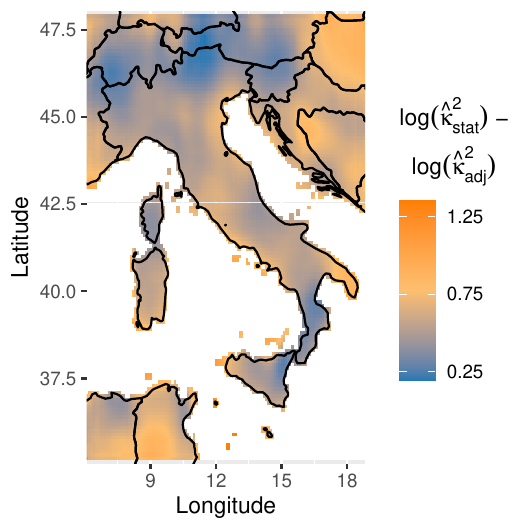}
         \label{fig:kappa2_space}
     \end{subfigure}
     \caption{Differences between the range parameters estimated by the stationary model and the adjusted non-stationary one, restricted to basis-function centers on land. Left: daily log-scale differences between the stationary range parameter and the spatial average of the nonstationary one. The dotted line indicates the mean difference of $0.8$. Right: annual average of the daily log-scale difference at each basis function center.}
     \label{fig:kappa2_t_s}
\end{figure}

The temporal correlation tends to be strong across all models, with estimated values of $\hat{\alpha}_t$ consistently exceeding $0.8$. Regarding spatial correlation, we analyze the changes in the range parameter between the stationary model and the adjusted non-stationary one. The comparison is made across both space and time, and is restricted to land cells where the adjustment occurs. From a temporal perspective, we compare the daily average of $\left\{ \widehat{\kappa}^2_{\mathrm{adj}}(\mathbf{u}) : w(\mathbf{u}) = 1 \right\}$ against the daily value of $\hat{\kappa}^2_{\mathrm{stat}}$. Figure \ref{fig:kappa2_t_s} shows the time series of their log-scale differences ($\log(\hat{\kappa}^2_{stat}) -\text{mean}_{\mathbf{u}}(\log(\widehat{\kappa}^2_\mathrm{adj}(\mathbf{u}))$), demonstrating that the nonstationary model exhibits a larger average correlation range on most days. This is likely due to the lack of flexibility permitted in the stationary model, causing it to artificially shrink the correlation range to avoid imposing dependence. From a spatial perspective, we compute the log-scale difference (i.e., $\log(\hat{\kappa}^2_{stat}) -\log(\widehat{\kappa}^2_\mathrm{adj}(\mathbf{u})$) at all basis-function centers on land, averaged over all 349 valid days, and visualize the result as a map in Figure~\ref{fig:kappa2_t_s}. The largest differences are concentrated in the Po Valley and along the coastline, where the nonstationary model captures the expected longer-range, anisotropic correlations that are characteristic of these areas. In contrast, both models produce similar correlation ranges in mountainous regions, where complex terrain shrinks the range.

\begin{figure}[htbp]
     \centering
     \includegraphics[width=\textwidth]{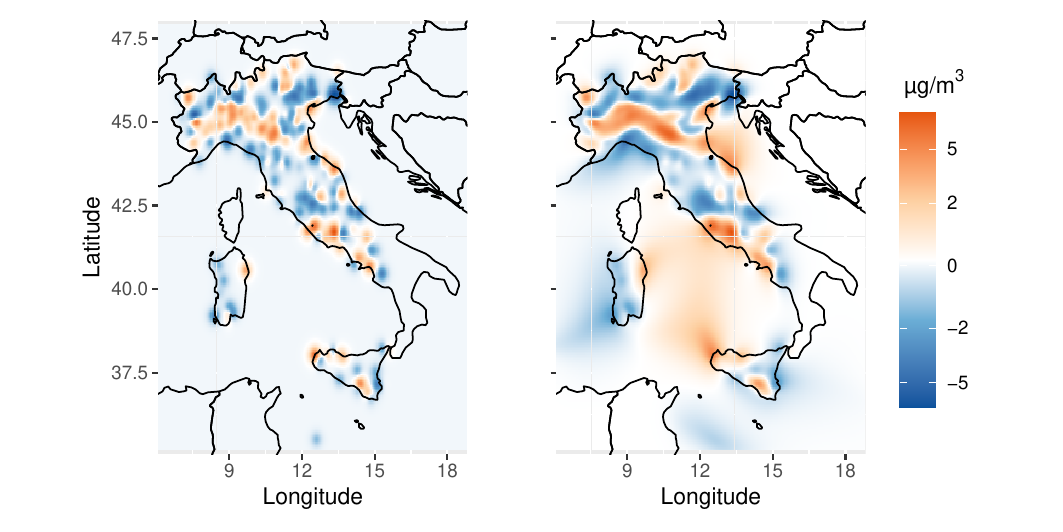}
     \caption{Estimated spatial component on November 1, 2023. Left: stationary model. Right: adjusted nonstationary model}
     \label{fig:sp_proc}
\end{figure}

To demonstrate the effect of different correlation parameters between the two models, Figure~\ref{fig:sp_proc} shows the estimated spatial components from both models for a representative day. The nonstationary model tends to produce a markedly smoother spatial component, which more coherently follows the underlying geographical and spatial patterns across the Italian peninsula. This is especially evident in the Po Valley, which appears as a more homogeneous region, in contrast to the localized “patchy” structures observed in the stationary model. Moreover, a larger correlation range in the sea induces a greater effect along the coast. 

\subsection{Cross-validation}

We also assess how well the different models reproduce observed NO$_2$ concentrations at monitoring locations via 10-fold cross-validation, performed separately for each of the $349$ eligible days. For a given day, the available stations are randomly partitioned into $10$ approximately equal folds. Each model is then iteratively fit to $90\%$ of the stations, and concentrations are predicted at the remaining $10\%$. We do not employ spatial stratification for the assignment of folds, as the set of active stations changes daily, and coverage can be highly uneven, which would result in imbalanced folds and unstable error estimates. Instead, we use random partitioning, ensuring each fold contains a comparable number of observations. Because the number of available stations varies by day, we report aggregated performance metrics for all days below in Table \ref{tab:station_cv_rmse}, with additional diagnostic figures provided in \ref{app:more_crossval}.

\begin{table}[htbp]
    \centering
    \caption{RMSE, 95\% Prediction Interval Coverage Probability (PICP), Mean 95\% Prediction Interval Width (MPIW). Arrows indicate the desirable direction.}
    \label{tab:station_cv_rmse}
    \begin{tabular}{lccc}
    \toprule
         Model & RMSE $\downarrow$ & PICP (95\%) & MPIW $\downarrow$ \\ 
         \midrule
         % LK\_def & & & \\
         Stationary & 6.544 &   94.030\% & 23.845 \\
         Nonstationary & 6.550 &  \textbf{94.037\% }&  24.038 \\
         Nonstationary (adjusted) & \textbf{6.530} & 94.022\% &  \textbf{23.826} \\
         \bottomrule
    \end{tabular}
\end{table}

As expected, all three spatial models perform near equivalently in aggregate, as station locations are frequently clustered together, reducing the ability to distinguish differences in model performance. This behavior is also consistent with existing literature on the effect of nonstationarity on predictive performance \citep{paciorek2006spatial}. Moreover, the stationary LatticeKrig model is already highly flexible with a large number of basis functions ($128 \times 128$), making the margin for improvement inherently small. Nevertheless, the correlation range refinement step systematically improves upon the original non-stationary model, and the adjusted non-stationary model achieves the lowest RMSE.

\subsection{Prediction to Fine Grids}

We use the adjusted nonstationary model to generate spatially complete, daily maps of NO$_2$ concentrations across all of Italy. For each of the 349 retained days, the fitted model is used to predict the latent daily mean field on a finer $0.05^{\circ}\times0.05^{\circ}$ grid ($255 \times 255$ pixels) over the same spatial domain (longitudes $6.1^{\circ}$--$18.8^{\circ}$E, latitudes $35.2^{\circ}$--$47.9^{\circ}$N). All covariates are aligned to the prediction grid via nearest-neighbor resampling, except for elevation, which is aggregated by averaging from its native higher resolution. This fine-grid prediction is made possible by embedding the STUN-estimated parameters within the LatticeKrig basis-function framework, rather than treating them as a strict GMRF on the observation grid. Because the basis functions $\Phi(\mathbf{s})$ can be evaluated at any spatial location $\mathbf{s}$, predictions are obtained on an arbitrarily fine target grid simply by evaluating $g(\mathbf{s})$ at the new locations, with no re-estimation of the covariance parameters. This decoupling of prediction resolution from estimation resolution is a direct consequence of extending LatticeVision to incorporate basis functions.

\begin{figure}[h]
\centering
\includegraphics[width=1.0\textwidth]{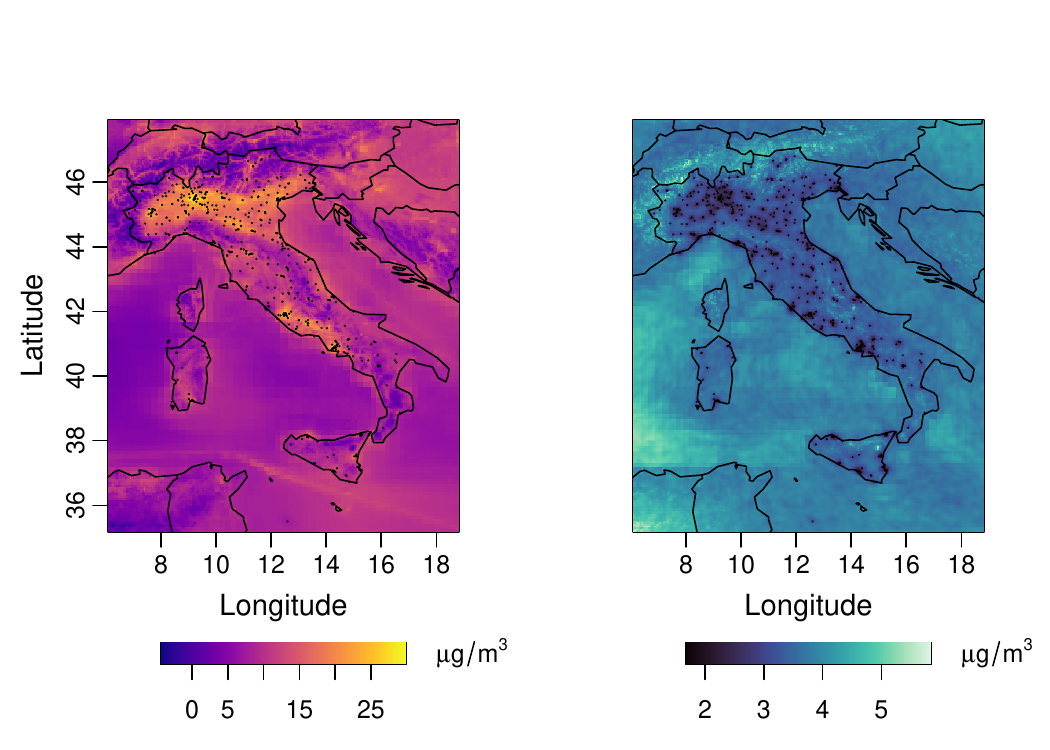}
\caption{ Fine-grid predictions for November 1, 2023. Left: posterior mean NO$_2$ concentration ($\mu$g/m$^3$) on the $0.05^{\circ}\times0.05^{\circ}$ prediction grid. Right: posterior standard error ($\mu$g/m$^3$) derived from a 1{,}000-member conditional simulation ensemble.}
\label{fig:cond_sim}
\end{figure}

To accompany the predictions with measures of uncertainty, we perform conditional simulation from the fitted Gaussian process. We generate 1{,}000 conditional draws, producing an ensemble of simulated fields that exactly respect both the fitted covariance structure and the observed station data. Pixelwise summaries of uncertainty, such as the posterior standard error, are then computed directly from this ensemble. The resulting predictions and standard errors for a representative day are shown in Figure~\ref{fig:cond_sim}. As expected, uncertainty is lowest in areas with dense station coverage and increases in topographically complex regions such as the Alps, rural areas with few nearby observations, and the sea.

It is important to note that the computational cost of prediction is modest. Generating the full fine-grid posterior mean for a single day requires 25 seconds on average, meaning the complete set of 349 daily maps can be produced in under three hours. Conditional simulation incurs a one-time required setup with the same computational cost as prediction (25 seconds), and each subsequent draw takes less than 2 seconds, allowing for very large ensembles to be generated. The resulting daily concentration maps and their associated uncertainty fields constitute a spatially complete, fine-resolution record of surface NO$_2$ over Italy, suitable for downstream applications in epidemiology, urban planning, and environmental policy assessment.

\section{Discussion}
\label{sec-discussion}

In this work, we introduce a spatial transfer-learning approach for combining sparse, irregular, point-level observational data with gridded, spatially complete numerical model outputs. The underlying spatial dependence structure is first estimated from the regularly gridded data and then transferred to the observation-level model. We apply this methodology to daily NO$_2$ concentrations over Italy, using monitoring-station observations and chemical transport model (CTM) outputs to produce spatially complete concentration maps on a fine prediction grid, while retaining an interpretable covariance model and providing rigorous uncertainty quantification.

Our approach builds on the LatticeKrig formulation, which represents a Gaussian process through compactly supported basis functions with coefficients governed by a sparse precision matrix, enabling scalable inference over large spatial domains. We address the change-of-support problem using two strategies. First, we expand amortized neural parameter estimation by explicitly incorporating the basis-function representation into a LatticeVision-style estimation pipeline. This allows the covariance structure that is learned from the regularly gridded CTM outputs to be transferred to point-level observations and subsequently projected onto a finer prediction grid. Second, we introduce a likelihood-based refinement step for the correlation range parameter when transitioning between areal and point-level data sources. In our application, we expect that the CTM outputs capture large-scale dependence and anisotropy but tend to smooth over fine-scale variability. Treating the CTM-derived anisotropy parameters as informative, we refine the local correlation range using station data, conditioning on the nonstationary, anisotropic correlation structure learned from the gridded data.

We apply this framework to both areal and point-level data. The CTM reconstruction experiment provides the most reliable way to evaluate predictions off of station locations. In this case, the nonstationary model achieves the best performance, with the strongest gains occurring during colder months. When applied to the point-level measurements, the adjusted non-stationary correlation range is larger than the stationary one, especially in the Po Valley and along coastlines. The different correlation structure is well demonstrated by the estimated spatial fields: the adjusted non-stationary process tends to show smoother behavior, which more coherently follows the underlying geographical and spatial patterns across the Italian peninsula. Moreover, the cross-validation scheme demonstrates the value of the correlation range refinement. Although cross-validation can be useful to assess predictive skill, in this application we feel it has limited value. This is due to the clustering of urban stations and the changing number of daily reporting stations.

Beyond point prediction, a key outcome of the proposed framework is the ability to generate spatially complete daily maps with uncertainty quantification. Using conditional simulation under the fitted nonstationary LatticeKrig model, we produce ensembles of plausible NO$_2$ fields on a $0.05^{\circ}\times0.05^{\circ}$ grid, enabling pixel-wise uncertainty summaries and uncertainty propagation for derived spatial quantities. The resulting uncertainty maps exhibit intuitive spatial patterns, with low uncertainty in regions of dense monitoring coverage and increased uncertainty in topographically complex or observation-sparse areas. Both predictions and conditional ensembles for uncertainty quantification incur minimal computational expense. 

Future research could explore several extensions to the framework proposed in this paper. First, while temporal dependence is incorporated in the mean through an ARX(1) specification, a more sophisticated handling of the temporal trend could reduce uncertainty and improve prediction on days with sparse station coverage. Second, although we focus on a single-resolution LatticeKrig model for clarity, this framework naturally extends to a multi-resolution construction. Including several resolutions results in covariance functions that more closely approximate standard families such as the Matérn. Third, the likelihood-based refinement step currently adjusts only the correlation range parameter; extending this refinement to additional parameters may improve predictive performance on observational data. Finally, more flexible weighting schemes in the refinement step based on station density, distance to observations, or terrain characteristics could be explored to improve localized adjustments.

\section{Acknowledgements}
\label{sec-ack}
We would like to thank Soutir Bandyopadhyay, Souvick Bera, and Adrienne Marshall for their insights and helpful discussions. This material is based upon work supported by the National Science Foundation Graduate Research Fellowship Program under Grant No. DGE-2137099. Any opinions, findings, and conclusions or recommendations expressed in this material are those of the authors and do not necessarily reflect the views of the National Science Foundation. 
% \begin{sloppypar}
% \textbf{Funder}: 
This project is co-funded under the National Recovery and Resilience Plan (NRRP), Mission 4 Component 2 Investment 1.3 - Call for tender No. 341 of 15/03/2022 of Italian Ministry of University and Research funded by the European Union – NextGenerationEU. Award Number: PE\_\allowbreak 00000018, Concession Decree No. 1558 of 11/10/2022 adopted by the Italian Ministry of University and Research, CUP F83C22001720001, GROWING RESILIENT INCLUSIVE AND SUSTAINABLE - GRINS \\
% \end{sloppypar}

\appendix
% \section{AQCLIM Details}

\section{Change of support}
\label{app:cosp}
The latent process is defined on the continuous domain $\mathcal{S} \subset \mathbb{R}^2$:
\begin{equation}
    y_t(\mathbf{s}) = \mathbf{x}_t(\mathbf{s})^{\top}\boldsymbol{\beta}_t + \alpha_t y_{t-1}(\mathbf{s}) + g_t(\mathbf{s}), \quad \mathbf{s} \in \mathcal{S},
\end{equation}
where $g_t(\mathbf{s}) \sim GP(0,C_{g_t}(\mathbf{s},\mathbf{s'}))$ and $C_{g_t}(\cdot,\cdot)$ is anisotropic and nonstationary. Observations from the monitoring stations are considered as realizations of the point-level Gaussian process $z_t^{(1)}(\cdot)$ defined as
\begin{equation}
    z_t^{(1)}(\mathbf{s}) \equiv y_t(\mathbf{s}) + \varepsilon^{(1)}_t(\mathbf{s}),
    \label{eq:staz_app}
\end{equation}
where $\varepsilon^{(1)}_t(\mathbf{s}) \sim \mathcal{N}(0, \tau_t^2), \tau_t^2 > 0$. Data from the chemical transport model are considered as realizations of an areal process representing imperfect measurements of areal averages of the latent process,
\begin{equation}
    z_{i,t}^{(2)} \equiv \bar{y}_{i,t} + \varepsilon^{(2)}_{i,t},
    \label{eq:ctm_app}
\end{equation}
where $\varepsilon^{(2)}_t \sim \mathcal{N}(0, \xi_t^2), \xi_t^2 > 0$ and $\bar{y}_{i,t}$ represents the exact areal averages of the latent process in $A_i \subset \mathcal{S}$ with $\bigcup_{i=1}^NA_i \equiv \mathcal{S}$. The collection $\{A_i\}_{i=1}^N$ induces a discretised spatial support $\mathcal{S}^d$ that represents the tiling of the domain. Every component of the latent process is also defined on this spatial support as
\begin{equation}
    \bar{y}_{i,t} = \frac{1}{|A_i|} \int_{A_i} y_t(\mathbf{s})\text{d}\mathbf{s}, 
\end{equation}

\begin{equation}
    \bar{\mathbf{x}}_{i,t} = \frac{1}{|A_i|} \int_{A_i} \mathbf{x}_t(\mathbf{s})\text{d}\mathbf{s}, 
\end{equation}

\begin{equation}
    \bar{g}_{i,t} = \frac{1}{|A_i|} \int_{A_i} g_t(\mathbf{s})\text{d}\mathbf{s}.
\end{equation}
The collection of these components on the limited set of tilings produces the vectors $\mathbf{\bar{y}_{t}} = (\bar{y}_{1,t},\ldots,\bar{y}_{N,t})^{\top}$ and $\mathbf{\bar{g}_{t}} = (\bar{g}_{1,t},\ldots,\bar{g}_{N,t})^{\top}$ where both $\mathbf{\bar{y}_{t}}$ and $\mathbf{\bar{g}_{t}} \in \mathbb{R}^N$. Covariates can be rearranged as a matrix $\overline{\mathbf{X}}_{t}= (\bar{\mathbf{x}}_{1,t},\ldots,\bar{\mathbf{x}}_{N,t}) \in \mathbb{R}^{p\times N}$ with $p$ covariates. 
Thus, the areal latent process can be completely defined on the discretised spatial support $\mathcal{S}^d$ as
\begin{equation}
    \mathbf{\bar{y}_{t}} = \overline{\mathbf{X}}_{t}^{\top}\boldsymbol{\beta}_t+ \alpha_t \mathbf{\bar{y}_{t-1}} + \mathbf{\bar{g}_{t}},
\end{equation}
where $\mathbf{\bar{g}_{t}} \sim \mathcal{N}(0,\Sigma_{\bar{g}_t})$. The elements of the matrix $\Sigma_{\bar{g}_t}$ correspond to the covariance between all pairs $\bar{g}_{i,t}$ and $\bar{g}_{j,t}$, calculated as
\begin{equation}(\Sigma_{\bar{g}_t})_{i,j}=COV(\bar{g}_{i,t},\bar{g}_{j,t}) = \frac{1}{|A_i||A_j|}\int_{A_i\times A_j}C_{g_t}(\mathbf{s},\mathbf{s}') \, d\mathbf{s} \,d\mathbf{s}'.
\end{equation}
The different correlation structure between $C_{g_t}(\mathbf{s},\mathbf{s}')$ and $\Sigma_{\bar{g}_t}$ depends on the spatial resolution of the tiling, as discussed below. Considering the $i$-th region $A_i$ and substituting in the areal model \eqref{eq:ctm_app}, the process of areal data becomes
\begin{equation}
    z_{i,t}^{(2)} = \bar{\mathbf{x}}_{i,t}^{\top}\boldsymbol{\beta}_t+ \alpha_t \bar{y}_{i,t-1} + \bar{g}_{i,t} + \varepsilon^{(2)}_{i,t},
\end{equation}
\begin{equation}
    z_{i,t}^{(2)} = \bar{\mathbf{x}}_{i,t}^{\top}\boldsymbol{\beta}_t+ \alpha_t (z_{i,t-1}^{(2)}- \varepsilon^{(2)}_{i,t-1}) + \bar{g}_{i,t} + \varepsilon^{(2)}_{i,t},
\end{equation}
\begin{equation}
    z_{i,t}^{(2)} = \bar{\mathbf{x}}_{i,t}^{\top}\boldsymbol{\beta}_t+ \alpha_t z_{i,t-1}^{(2)} + \bar{g}_{i,t} + \varepsilon^{(2)}_{i,t} - \alpha_t\varepsilon^{(2)}_{i,t-1},
\end{equation}
\begin{equation}
    z_{i,t}^{(2)} = \bar{\mathbf{x}}_{i,t}^{\top}\boldsymbol{\beta}_t+ \alpha_t z_{i,t-1}^{(2)} + \eta^{(2)}_{i,t},
\end{equation}
with
\begin{equation}
    \eta^{(2)}_{i,t} = \bar{g}_{i,t} + \varepsilon_{i,t}^{(2)} - \alpha_t\varepsilon_{i,t-1}^{(2)}.
\end{equation}
Here,  $\eta^{(2)}_{i,t}$ captures the spatial component and the error term, in particular the collection $\bm{\eta}^{(2)}_{t} = (\eta^{(2)}_{1,t},\ldots,\eta^{(2)}_{N,t})^{\top} \sim \mathcal{N}(0,\Sigma_{\eta_t}^{(2)})$ and $\Sigma_{\eta_t}^{(2)} = \Sigma_{\bar{g}_t}+I(\alpha_t^2\xi^2_{t-1} + \xi^2_t))$. From $\bm{\eta}^{(2)}_{t}$, we use the STUN estimator to retrieve estimates of parameters for the covariance matrix $\Sigma_{\bar{g}_t}$. These parameters are then used to inform the point-level model (Eq. \ref{eq:staz_app}). Despite the temporal correlation in the error term being induced by the first-order moving average, this is not modeled explicitly, as the analysis is performed one day at a time. Moreover, the conditional variance $\operatorname{VAR}(\bm{\eta}^{(2)}_{t} \mid \bm{\eta}^{(2)}_{t-1})$ is less than or equal to the marginal variance; therefore, considering the marginal variance does not introduce substantial error. Regarding the connection between  $\Sigma_{\bar{g}_t}$ and  $C_{g_t}$, for sufficiently small areal units $|A_i|$, the induced covariance $\Sigma_{\bar{g}_t}$ closely approximates the point-level covariance $C_{g_t}$. However, as the support size increases, $\Sigma_{\bar{g}_t}$ progressively smooths out the short-range dependence structure. Therefore, when moving from areal data to point-level data, we include a term that captures the local variability that is smoothed-out by the averages. This term is assumed to follow a Gaussian process with short-range spatial dependence. Specifically, the areal averages $\bar{y}_{i,t}$ and the point process $y_t(\mathbf{s})$ can be related as
\begin{equation}
    y_t(\mathbf{s})=\bar{y}_{i,t} + \psi_t(\mathbf{s}), \qquad \mathbf{s} \in A_i,
    \label{eq:cosp1}
\end{equation}
where $\psi_t(\mathbf{s}) \sim GP(0,C_{\psi_t}(\mathbf{s},\mathbf{s'}))$ and $C_{\psi_t}$ is a valid covariance function where $C_{\psi_t} = f(C_{g_t})$. We now derive an explicit connection between the station measurements 
$z_{t}^{(1)}(\mathbf{s})$ and the CTM areal data $z_{i,t}^{(2)}$. Substituting Eq.~\ref{eq:cosp1} into the point-level model (Eq.~\ref{eq:staz_app}), where $i$ indexes the grid cell $A_i$ containing location $\mathbf{s}$, gives
\begin{equation}
    z_t^{(1)}(\mathbf{s}) = \bar{y}_{i,t} + \psi_t(\mathbf{s}) + \varepsilon_t(\mathbf{s}), \qquad \mathbf{s} \in \mathcal{S}
\end{equation}
\begin{equation}
    z_t^{(1)}(\mathbf{s}) = \bar{\mathbf{x}}_{i,t}^{\top}\boldsymbol{\beta}_t + \alpha_t \bar{y}_{i,t-1} + \bar{g}_{i,t} + \psi_t(\mathbf{s}) + \varepsilon^{(1)}_t(\mathbf{s}),
\end{equation}
\begin{equation}
    z_t^{(1)}(\mathbf{s}) = \bar{\mathbf{x}}_{i,t}^{\top}\boldsymbol{\beta}_t + \alpha_t (z_{i,t-1}^{(2)}- \varepsilon^{(2)}_{i,t-1}) + \bar{g}_{i,t} + \psi_t(\mathbf{s}) + \varepsilon^{(1)}_t(\mathbf{s}),
\end{equation}
\begin{equation}
    z_t^{(1)}(\mathbf{s}) = \bar{\mathbf{x}}_{i,t}^{\top}\boldsymbol{\beta}_t + \alpha_t z_{i,t-1}^{(2)} + \bar{g}_{i,t} - \alpha_t\varepsilon^{(2)}_{i,t-1} + \psi_t(\mathbf{s}) + \varepsilon^{(1)}_t(\mathbf{s}).
\end{equation}
For ease of exposition, let
\begin{equation}
    \eta^{(1)}_{t}(\mathbf{s}) = \bar{g}_{i,t} +  \varepsilon^{(1)}_t(\mathbf{s}) - \alpha_t\varepsilon^{(2)}_{i,t-1} + \psi_t(\mathbf{s}),
\end{equation}
where $\eta^{(1)}_{t}(\mathbf{s}) \sim GP(0,C_{\eta_t}^{(1)})$. The covariance function $C_{\eta_t}^{(1)}(\mathbf{s},\mathbf{s}')$ takes different forms depending on the relative locations of $\mathbf{s}$ and $\mathbf{s}'$: 
\begin{itemize}
    \item if $\mathbf{s}=\mathbf{s}' \in A_i$,
    \begin{equation}
    C_{\eta_t}^{(1)}(\mathbf{s},\mathbf{s}) = (\Sigma_{\bar{g}_t})_{i,i} + \alpha^2_t\xi^2_{t-1} + C_{\psi_t}(\mathbf{s},\mathbf{s}) + \tau_t^2, 
\end{equation}
\item if $\mathbf{s} \neq \mathbf{s}' \in A_i$,
\begin{equation}
    C_{\eta_t}^{(1)}(\mathbf{s},\mathbf{s}') = (\Sigma_{\bar{g}_t})_{i,i} + \alpha^2_t\xi^2_{t-1} + C_{\psi_t}(\mathbf{s},\mathbf{s}'),
\end{equation}
\item if $\mathbf{s} \in A_i$ and $\mathbf{s}' \in A_j$,
\begin{equation}
    C_{\eta_t}^{(1)}(\mathbf{s},\mathbf{s}') = (\Sigma_{\bar{g}_t})_{i,j} + C_{\psi_t}(\mathbf{s},\mathbf{s}'),
\end{equation}
\end{itemize}
where $(\Sigma_{\bar{g}_t})_{i,j}$ indicates the $i,j$-th element of the matrix $\Sigma_{\bar{g}_t}$. Introducing the term $\eta^{(1)}_t(\mathbf{s})$ allows us to write the point-level model as
\begin{equation}
    z_t^{(1)}(\mathbf{s}) = \bar{\mathbf{x}}_{i,t}^{\top}\beta + \alpha_t z_{i,t-1}^{(2)} + \eta^{(1)}_t(\mathbf{s}).
\end{equation}

To summarize, both the point-level covariance function $C_{\eta_t}^{(1)}(\cdot,\cdot)$ and the areal covariance matrix $\Sigma_{\eta_t}^{(2)}$ encode the same large-scale  spatial structure determined by $\Sigma_{\bar{g}_t}$. For sufficiently small areal units $|A_i|$, the measurement errors are approximately equal ($\tau^2_t \simeq \xi^2_t$) and the sub-grid component $\psi_t(\mathbf{s})$ is negligible ($C_{\psi_t}(\cdot,\cdot)\simeq 0$), so that $C_{\eta_t}^{(1)}(\cdot,\cdot)$ and $\Sigma_{\eta_t}^{(2)}$ are almost equivalent. In the case of larger $|A_i|$, as is assumed in our application, $\Sigma_{\eta_t}^{(2)}$ does not include the short-range structure from $C_{\psi_t}$. We correct for this by introducing an additional range parameter estimated from the station data, which has the overall effect of reducing the correlation range.

\section{Neural Estimator Details}
\label{sec:appendix_STUN}

The STUN (Spatial TransUNet) neural estimator was first introduced in \cite{LatticeVision}, and is based on the TransUNet architecture of \citet{chen2021transunet}. STUN combines a symmetric, UNet-style encoder–decoder with a Vision Transformer bottleneck. The encoder (respectively decoder) consists of a series of convolutional blocks that progressively downsample (respectively upsample) the spatial resolution, while the bottleneck applies multi-head self-attention across patches of the image to capture long-range dependencies. Skip connections between corresponding encoder and decoder layers preserve fine-scale spatial detail throughout the network. 

The network is trained with a batch size of $b=64$, so each training input has shape $[b,r,h_A,w_A]=[64,30,128,128]$, with corresponding target parameter images of shape $[64,3,128,128]$. We use the AdamW optimizer \citep{adamw} with a step-wise learning rate schedule and early stopping triggered after 10 epochs without improvement on the validation set, training for a maximum of 200 epochs. The loss function is mean squared error (MSE), computed on standardized parameter values within the training loop to avoid scale imbalance across the three parameter channels. Data augmentation is restricted to spatial translation and random field negation, as other common augmentation techniques (such as rotation) often do not preserve the statistical properties of the image. 

STUN is implemented in PyTorch \citep{paszke2019pytorch} and trained on a single NVIDIA RTX A6000 GPU. Synthetic data generation is performed in \texttt{R} on a laptop with a 2.20 GHz Intel Core i9-14900HX processor and 32 GB RAM, using the \texttt{LatticeKrig} package \citep{LKrigpackage} to simulate spatial fields. The complete synthetic dataset---comprising 20,000 training pairs, each with $r=30$ replicates---required approximately 16 hours to generate and occupies 80 GB of storage. Data are compressed and stored in the \texttt{HDF5} format to allow easy access using both \texttt{R} and \texttt{Python}.

\section{Additional Cross Validation Results}
\label{app:more_crossval}

\begin{figure}[htbp]
\centering
\includegraphics[width=1.0\textwidth]{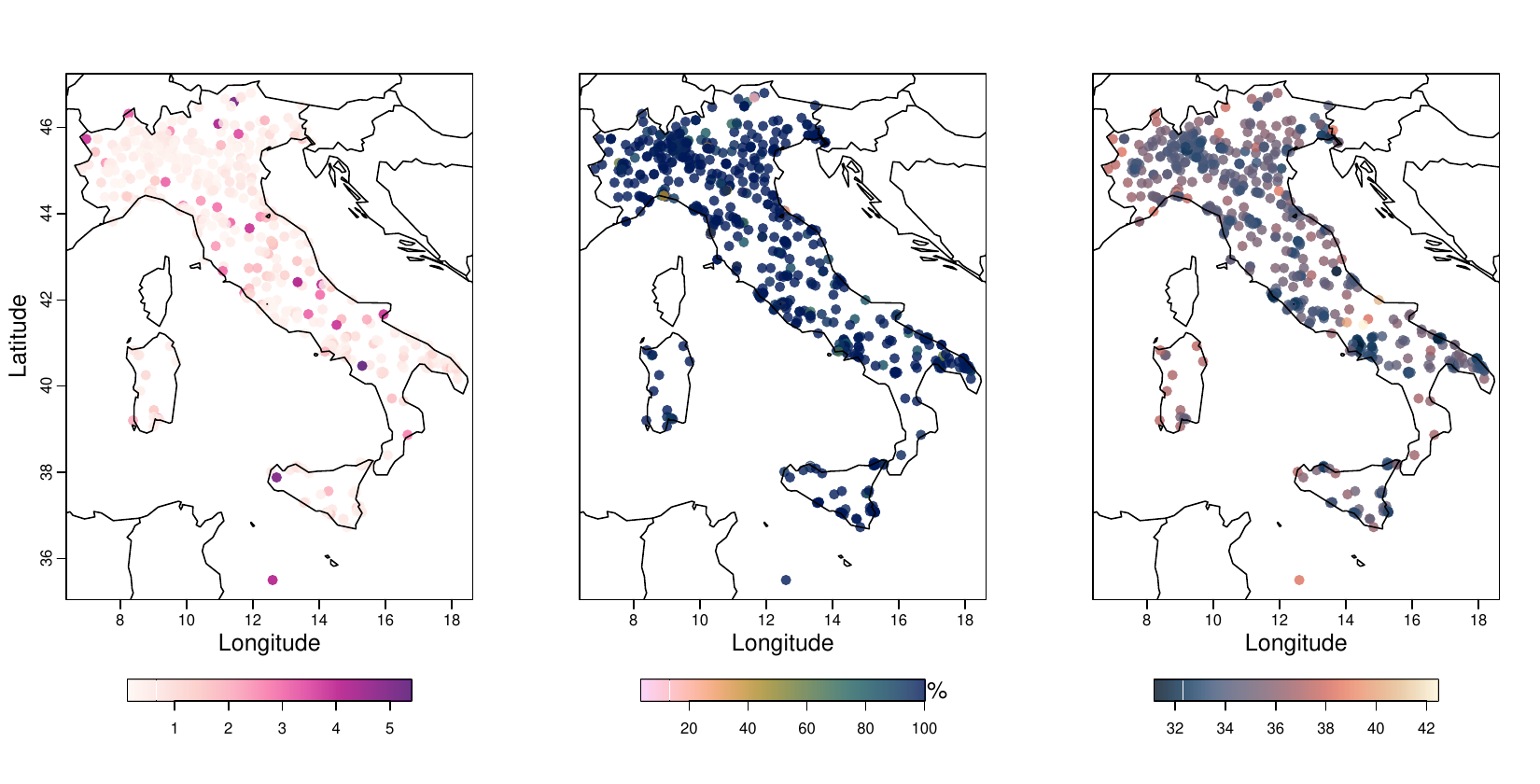}
\caption{Station-level cross-validation metrics for the nonstationary 
(adjusted) model, aggregated over all 349 days. From left to right: RMSE weighted by the local annual mean NO$_2$ concentration, 95\% prediction interval coverage probability, and mean prediction 
interval width ($\mu$g/m$^3$).}
\label{fig:station_bubbleplot}
\end{figure}

To further examine the reliability of the nonstationary model, Figure~\ref{fig:station_bubbleplot} shows the station-wise spatial distribution of weighted RMSE, coverage, and prediction interval widths. The coverage is well-calibrated and spatially homogeneous, lending confidence to the nominal 95\% intervals despite the large number of days and stations contributing to the aggregate PICP. Prediction interval widths and RMSE are consistent with the complexity of terrain and regions with fewer stations. Overall, predictive performance is strong across the entire domain, with the exception of a small number of outlier stations situated in atypical environments, where local emission sources (such as road tunnels) produce concentration patterns that are not well represented by the correlation structure of the model. 

\bibliographystyle{elsarticle-harv}
\bibliography{refs}

\end{document}